\documentclass[12pt]{article}
\usepackage{a4wide}
\usepackage{latexsym}
\usepackage{amsmath}
\usepackage{amsfonts}
\usepackage{amscd}
\usepackage{cite}

\usepackage{pslatex}
\usepackage{graphicx}
\usepackage[latin1]{inputenc}
\usepackage[T1]{fontenc}

\allowdisplaybreaks

\newcommand{\bq}{\begin{eqnarray}}
\newcommand{\eq}{\end{eqnarray}}
\newcommand{\eps}{\varepsilon}

\begin{document}

\thispagestyle{empty}

\begin{flushright}
 MaPhy-AvH/2017-05
 \\
 MITP/17-038
\end{flushright}

\vspace{1.5cm}

\begin{center}
  {\Large\bf Analytic continuation and numerical evaluation of the kite integral and the equal mass sunrise integral \\
  }
  \vspace{1cm}
  {\large Christian Bogner ${}^{a}$, Armin Schweitzer ${}^{a}$ and Stefan Weinzierl ${}^{b}$ \\
  \vspace{1cm}
      {\small ${}^{a}$ \em Institut f{\"u}r Physik, Humboldt-Universit{\"a}t zu Berlin,}\\
      {\small \em D - 10099 Berlin, Germany}\\
  \vspace{2mm}
      {\small ${}^{b}$ \em PRISMA Cluster of Excellence, Institut f{\"u}r Physik, }\\
      {\small \em Johannes Gutenberg-Universit{\"a}t Mainz,}\\
      {\small \em D - 55099 Mainz, Germany}\\
  } 
\end{center}

\vspace{2cm}

\begin{abstract}\noindent
  {
We study the analytic continuation of Feynman integrals from the kite family, expressed
in terms of elliptic generalisations of (multiple) polylogarithms.
Expressed in this way, the Feynman integrals are functions of two periods of an elliptic curve.
We show that all what is required is just the analytic continuation of these two periods.
We present an explicit formula for the two periods for all values of $t \in {\mathbb R}$.
Furthermore, the nome $q$ of the elliptic curve satisfies over the complete range in $t$
the inequality $|q|\le 1$, where $|q|=1$ is attained only at the singular points $t\in\{m^2,9m^2,\infty\}$.
This ensures the convergence of the $q$-series expansion of the $\mathrm{ELi}$-functions
and provides a fast and efficient evaluation of these Feynman integrals.
   }
\end{abstract}

\vspace*{\fill}

\newpage

\section{Introduction}
\label{sec:intro}

Precision calculations in high-energy particle physics require the evaluation of Feynman loop integrals.
Within an analytical approach towards Feynman loop integrals one
computes first the Feynman integrals in terms of a specific class of 
transcendental function in a particular kinematic region (usually the Euclidean region).
In a second step one studies the analytic continuation into the full kinematic region and the numerical
evaluation of the transcendental functions.
This procedure has been successful for a wide class of Feynman integrals, which evaluate
to multiple polylogarithms \cite{Goncharov_no_note,Goncharov:2001,Borwein,Moch:2001zr}.
Numerical algorithms are available to evaluate multiple polylogarithms
for arbitrary complex-valued 
arguments \cite{Vollinga:2004sn,Gehrmann:2001pz,Gehrmann:2001jv,Maitre:2005uu,Maitre:2007kp,Frellesvig:2016ske}.

It is well-known that not all Feynman integrals can be expressed in terms of multiple polylogarithms.
The simplest counter-example is given by the two-loop equal mass sunrise integral.
By now, this integral has been studied extensively 
in the literature \cite{Broadhurst:1993mw,Berends:1993ee,Bauberger:1994nk,Bauberger:1994by,Bauberger:1994hx,Caffo:1998du,Laporta:2004rb,Kniehl:2005bc,Groote:2005ay,Groote:2012pa,Bailey:2008ib,MullerStach:2011ru,Adams:2013nia,Bloch:2013tra,Adams:2014vja,Adams:2015gva,Adams:2015ydq,Bloch:2016izu,Adams:2017ejb,Remiddi:2013joa}.
This is of course justified by the fact that the sunrise integral is the ``first'' Feynman integral which cannot be expressed
in terms of multiple polylogarithms.
Further examples of Feynman integrals not expressible in terms of multiple polylogarithms are discussed 
in \cite{Sabry:1962,Remiddi:2016gno,Adams:2016xah,Bloch:2014qca,Primo:2017ipr,Bonciani:2016qxi,vonManteuffel:2017hms}.
In this paper we are interested in a larger family of Feynman integrals, 
the family of kite integrals \cite{Sabry:1962,Remiddi:2016gno,Adams:2016xah}.
By the term ``family of Feynman integrals'' we mean Feynman integrals which differ only in the powers of their propagators,
including the case, where one or more exponents are zero.
In this case the corresponding propagator is absent and the Feynman integral reduces to a sub-topology.
The kite integral contains as a sub-topology the equal-mass sunrise integral.
Like the sunrise integral, the kite integral cannot be expressed in terms of multiple polylogarithms.
Both the kite integral and the sunrise integral are two-point functions, depending on a variable $t=p^2$ equal
to the incoming (or outgoing) momentum squared.
In two recent publications we showed that the equal-mass sunrise integral \cite{Adams:2015ydq} 
and the kite integral \cite{Adams:2016xah} can be expressed in a neighbourhood of $t=0$ 
to all orders in the dimensional regularisation parameter $\eps$
in terms of elliptic generalisations of (multiple) polylogarithms.
These generalisations are denoted as $\mathrm{ELi}$-functions and are functions of several variables.
Of particular interest is the dependence on one particular variable denoted by $q$.
Feynman integrals are related to periods of algebraic varieties \cite{Bloch:2005,Bogner:2007mn}.
In the case of the family of kite integrals the non-trivial algebraic variety which prohibits an evaluation in terms
of multiple polylogarithms is an elliptic curve.
The variable $q$ is the nome of the elliptic curve.
By the modularity theorem, every elliptic curve over ${\mathbb Q}$ has a modular parametrisation
and in \cite{Adams:2017ejb} one of the authors showed that one may express the family of kite integrals 
to all orders in the dimensional regularisation parameter $\eps$ as iterated integrals of modular forms.
The $\mathrm{ELi}$-functions give the $q$-series expansion of these iterated integrals of modular forms.
At $t=0$ we have $q=0$ and therefore $|q|<1$ in a neighbourhood of $t=0$.
In this neighbourhood the $\mathrm{ELi}$-functions provide a simple and convenient evaluation
of the Feynman integrals of the kite family.

In this paper we consider the analytic continuation and the numerical evaluation of the integrals of the kite family
in the complete kinematic region $t \in {\mathbb R}$.
We note that in a recent paper \cite{Passarino:2017EPJC}
the analytic continuation of the $\mathrm{ELi}$-functions has been discussed.
It is one of the results of the present paper, that for the family of kite integrals the analytic continuation 
of the $\mathrm{ELi}$-functions is not needed.
The nome $q$ is defined in terms of the two periods $\psi_1$ and $\psi_2$ 
of the elliptic curve by $q=\exp(i \pi \psi_2/\psi_1)$.
All what is needed is the analytic continuation of the two periods $\psi_1$ and $\psi_2$.
We present an explicit formula for the two periods for all values of $t \in {\mathbb R}$.
It turns out that once the two periods $\psi_1$ and $\psi_2$ are properly defined, we have
$|q| \le 1$ for all $t \in {\mathbb R}$.
and $|q|=1$ is attained only at the singular points $t\in\{m^2,9m^2,\infty\}$.
As a second main result of this paper we show that the same expressions in terms of the $\mathrm{ELi}$-functions which we found 
for the Feynman integrals of the kite family in the region around $t=0$ hold for all $t \in {\mathbb R}$.
The $\mathrm{ELi}$-functions provide therefore a fast and efficient way to evaluate these integrals
over the complete kinematic range.

Of course, Feynman integrals may also be evaluated 
by purely numerical methods \cite{Nagy:2003qn,Anastasiou:2007qb,Gong:2008ww,Becker:2010ng,Becker:2012aq,Becker:2012nk,Becker:2012bi,Catani:2008xa,Hernandez-Pinto:2015ysa,Buchta:2015wna,Sborlini:2016gbr}.
A flexible numerical method to check Feynman integrals at individual kinematic points is given by
sector decomposition \cite{Binoth:2000ps,Bogner:2007cr,Smirnov:2008aw}.
We compare numerical results from the evaluation of the Feynman integrals from the kite family expressed
in terms of $\mathrm{ELi}$-functions
with numerical results from the {\tt SeCDec} program \cite{Borowka:2015mxa,Borowka:2012yc,Borowka:2013cma,Borowka:2014aaa}.
We find perfect agreement in all kinematic regions, including the regions close to the thresholds.

As already mentioned, we only need to continue analytically the two periods $\psi_1$ and $\psi_2$.
Taking Feynman's $i0$-prescription into account, the two periods are continuous functions of $t$.
We express them in terms of complete elliptic integrals of the first kind.
The complete elliptic integral of the first kind $K(z)$ has a branch cut in the variable $z^2$ at $[1,\infty[$.
If we cross this branch cut, we have to compensate the discontinuity of $K(z)$ by taking into account
the monodromy around $z^2=1$.

This paper is organised as follows:
In section~\ref{sec:convention} we briefly recall the conventions as arguments of functions approach a branch cut
for mathematical software on the one hand
and Feynman's $i0$-prescription in physics on the other hand.
In section~\ref{sec:elliptic_curve} we study a family of elliptic curves $E_t$ and their periods.
We express the periods in terms of complete elliptic integrals of the first kind, such that the
periods are continuous functions as $t$ varies continuously.
Equipped with the appropriate definition of the periods we discuss the analytic
continuation of the Feynman integrals of the kite family in section~\ref{sec:feynman_integrals}.
The correct definition of the periods requires to take a monodromy matrix at $t=m^2$ into account.
Section~\ref{sect:picard_lefschetz} is devoted to the derivation of this monodromy matrix.
In section~\ref{sect:numerical_results} we show numerical results for three Feynman integrals
from the kite family.
Finally, our conclusions are given in section~\ref{sect:conclusions}.
Appendix~\ref{sec:convention_roots} gives information on our conventions regarding the
roots of the cubic polynomial of the Weierstrass normal form.
In appendix~\ref{sec:ELi} we summarise the definition of the $\mathrm{ELi}$-functions and 
the definition of the $\overline{\mathrm{E}}$-functions, 
the latter being linear combinations of the former.
Appendix~\ref{sec:agm} reviews the algorithm for the numerical computation of the complete
elliptic integrals of the first kind based on the arithmetic-geometric mean.

\section{Conventions}
\label{sec:convention}

We will encounter mathematical functions of a complex variable $z$, like $\sqrt{z}$ or the complete elliptic integral $K(z)$.
These functions have branch cuts.
Let us clarify, which values to assign to these functions on the branch cuts.
The standard convention for mathematical software is as follows:
{\em Implementations shall map a cut so the function is continuous as the cut is approached coming
around the finite endpoint of the cut in a counter clockwise direction} \cite{C99standard}.  
With this convention, cuts on the positive real axis are 
continuous to the lower complex half-plane while cuts on the negative real axis are continuous to the upper complex half-plane.

However, this is not always what we want in physics.
In physics the analytic continuation is dictated by Feynman's $i0$-prescription, where we substitute 
a (real) variable $t$ by $t \rightarrow t + i0$.
The symbol $+i0$ denotes an infinitesimal positive imaginary part.
The small imaginary part overrides the mathematical convention above.

To give an example we have
\bq
 \sqrt{-t} 
 \;\; = \;\;
 \left\{
  \begin{array}{rl}
    i \sqrt{\left|t\right|}, & t > 0, \\
    \sqrt{\left|t\right|}, & t \le 0, \\
  \end{array}
 \right.
 & &
 \sqrt{-t-i0} 
 \;\; = \;\;
 \left\{
  \begin{array}{rl}
    -i \sqrt{\left|t\right|}, & t > 0, \\
    \sqrt{\left|t\right|}, & t \le 0. \\
  \end{array}
 \right.
\eq
In the sequel we will always assume that a small positive imaginary part is added to the variable $t$.

Furthermore we will deal with equations of the form
\bq
 y^2 & = & \left(x-e_1\right) \left(x-e_2\right) \left(x-e_3\right) \left(x-e_4\right),
\eq
and we would like to express $y$ as the square root of the right-hand side.
The square root of the right-hand side may be viewed as a multi-valued function of $x$, taking two possible value which differ by a sign.
We are interested in a single-valued (and continuous) function of $x$.
Of course this cannot be done on the entire complex plane ${\mathbb C}$, but only on the complex plane minus some cuts.
Let us first assume that the roots $e_1$, $e_2$, $e_3$ and $e_4$ are real and ordered as
\bq
 e_1 \; < \; e_2 \; < \; e_3 \; < \; e_4.
\eq
Let us choose the cuts to be given by the line segment from $e_1$ to $e_2$ and by the line segment from $e_3$ to $e_4$.
Then we may choose $y$ as a single-valued and continuous function on ${\mathbb C} \backslash ([e_1,e_2] \cup [e_3,e_4])$.
It is not too difficult to show that one possible choice is given by
\bq
 y & = & \sqrt{x-e_1} \sqrt{x-e_2} \sqrt{x-e_3} \sqrt{x-e_4},
\eq
the other choice is given by
\bq
 y & = & - \sqrt{x-e_1} \sqrt{x-e_2} \sqrt{x-e_3} \sqrt{x-e_4}.
\eq
These functions have branch cuts at 
\bq
 e_i + \lambda, 
 & & 
 \lambda \in {\mathbb R}_{\le 0}.
\eq
On the interval $[e_2,e_3]$ we have a superposition of the branch cuts starting at $e_3$ and $e_4$, making the function continuous there.
The same happens on the interval $]-\infty,e_1]$, where we have a superposition of all four branch cuts.
It is worth noting that the naive guess
\bq
\label{wrong_choice}
 y & = & \sqrt{\left(x-e_1\right) \left(x-e_2\right) \left(x-e_3\right) \left(x-e_4\right)}
\eq
does in general not give a single-valued continuous function of $x$ on ${\mathbb C} \backslash ([e_1,e_2] \cup [e_3,e_4])$.
The cuts of eq.~(\ref{wrong_choice}) are in general algebraic functions of $x$ and determined by
\bq
 \left(x-e_1\right) \left(x-e_2\right) \left(x-e_3\right) \left(x-e_4\right) - \lambda \; = \; 0,
 & & 
 \lambda \in {\mathbb R}_{\le 0}.
\eq
Finally, let us remove the assumption $e_i \in {\mathbb R}$ and let us consider the general case $e_i \in {\mathbb C}$. 
We continue to denote by $[e_i,e_j]$ the line segment from $e_i$ to $e_j$, now in the complex plane.
Then we may express $y$ as a single-valued and continuous function on ${\mathbb C} \backslash ([e_1,e_2] \cup [e_3,e_4])$ through
\bq
 y & = &
 \pm
 \left(e_2-e_1\right) \left(e_4-e_3\right)
 \sqrt{\frac{x-e_1}{e_2-e_1}}
 \sqrt{\frac{x-e_2}{e_2-e_1}}
 \sqrt{\frac{x-e_3}{e_4-e_3}}
 \sqrt{\frac{x-e_4}{e_4-e_3}}.
\eq

\section{The elliptic curve}
\label{sec:elliptic_curve}

We consider a family of elliptic curves $E_t \subset {\mathbb P}^2({\mathbb C})$ given in the chart $z=1$ by the Weierstrass normal form
\bq
\label{WNF_with_g2_g3}
 y^2 & = & 4 x^3 - g_2(t) x - g_3(t),
\eq
with
\bq
 g_2(t)
 & = &
 \frac{1}{12\mu^8} \left(3m^2-t\right) \left(3m^6 - 3 m^4 t + 9 m^2 t^2 - t^3\right),
 \nonumber \\
 g_3(t)
 & = &
 \frac{1}{216 \mu^{12}} \left( 3 m^4 + 6 m^2 t - t^2 \right) \left( 9 m^8 - 36 m^6 t + 30 m^4 t^2 - 12 m^2 t^3 + t^4\right).
\eq
As discussed previously in the literature, this elliptic curve is associated with 
the equal-mass sunrise integral \cite{MullerStach:2011ru,Adams:2013nia,Bloch:2013tra,Adams:2014vja,Adams:2015gva,Adams:2015ydq,Bloch:2016izu,Adams:2017ejb}. Eq.~(\ref{WNF_with_g2_g3}) is obtained by transforming the equation
\bq
 {\mathcal F} & = & 0,
\eq
where
\bq
 {\mathcal F} & = & - x_1 x_2 x_3 t + m^2 \left( x_1 + x_2 + x_3 \right) \left( x_1 x_2 + x_2 x_3 + x_3 x_1 \right).
\eq
denotes the second graph polynomial 
of the equal-mass sunrise integral to the Weierstrass normal form.
Graph polynomials are reviewed in \cite{Bogner:2010kv}.
We may factor the cubic polynomial on the right-hand side of eq.~(\ref{WNF_with_g2_g3}). We then obtain
\bq
 y^2 = 4 \left(x-e_1\right)\left(x-e_2\right)\left(x-e_3\right),
 \;\;\;\;\;\;
 \mbox{with} 
 \;\;\;
 e_1+e_2+e_3=0,
\eq
where the roots are given by
\bq
\label{def_roots}
 e_1 
 & = &
 \frac{1}{24 \mu^4} \left( -t^2 + 6 m^2 t + 3 m^4 + 3 \left( m^2 - t\right)^{\frac{3}{2}} \left( 9 m^2 - t \right)^{\frac{1}{2}} \right),
 \nonumber \\
 e_2 
 & = &
 \frac{1}{24 \mu^4} \left( -t^2 + 6 m^2 t + 3 m^4 - 3 \left( m^2 - t\right)^{\frac{3}{2}} \left( 9 m^2 - t \right)^{\frac{1}{2}} \right),
 \nonumber \\
 e_3 
 & = &
 \frac{1}{24 \mu^4} \left( 2 t^2 - 12 m^2 t - 6 m^4 \right).
\eq
In appendix~\ref{sec:convention_roots} we discuss the motivation for defining the roots as in eq.~(\ref{def_roots}).

The roots $e_1$, $e_2$ and $e_3$ vary with $t$.
\begin{figure}
\begin{center}
\includegraphics[scale=1.0]{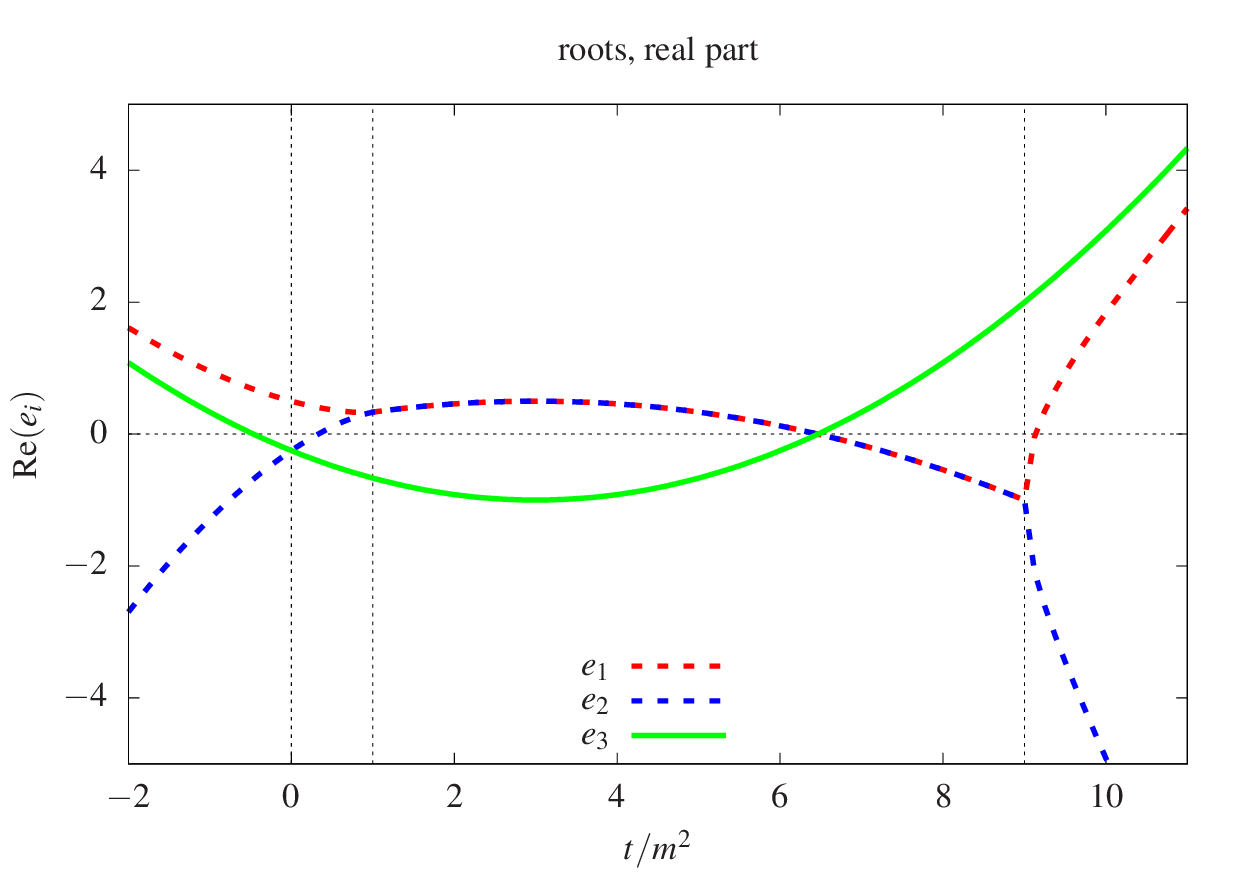}
\includegraphics[scale=1.0]{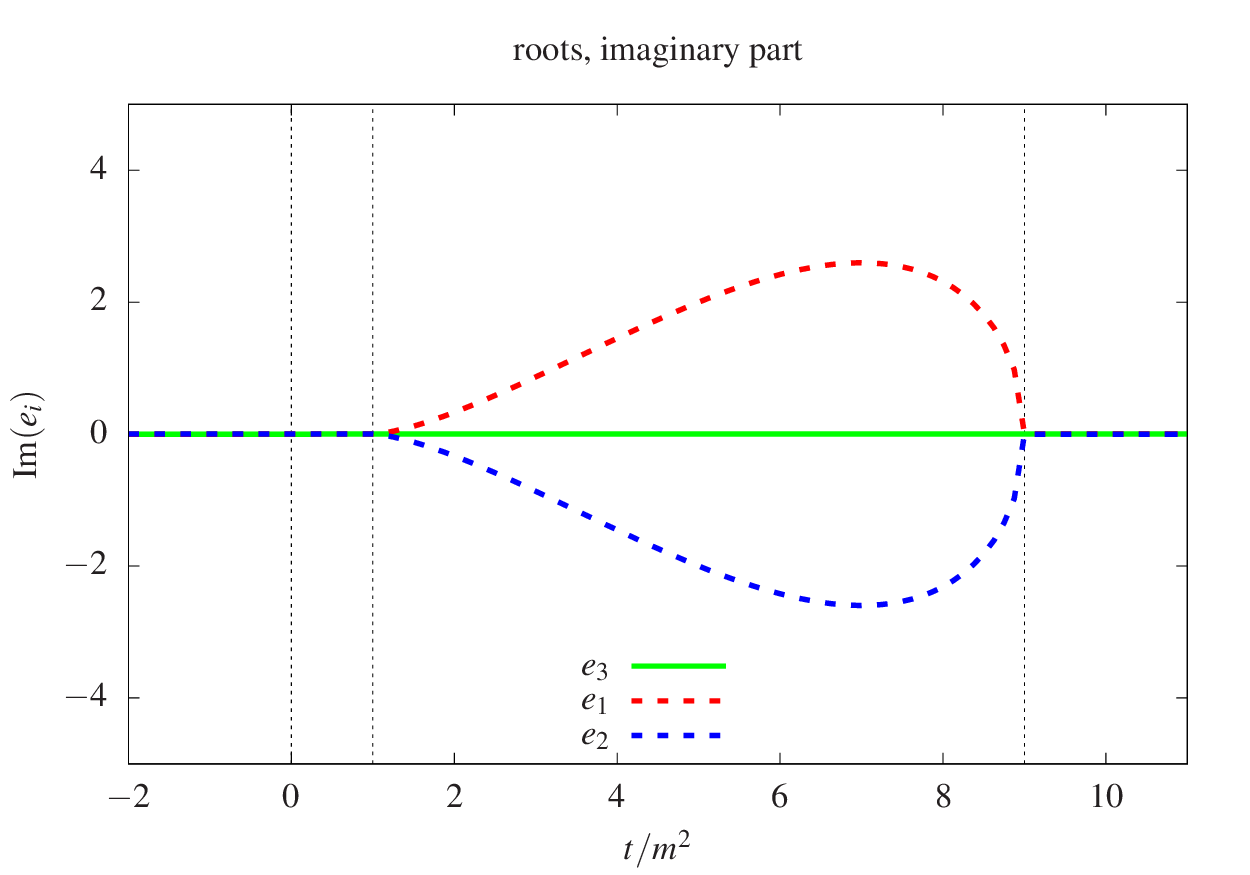}
\end{center}
\caption{\label{fig_roots}
The variation of the roots $e_1$, $e_2$ and $e_3$ of the elliptic curve $y^2 = 4 (x-e_1)(x-e_2)(x-e_3)$ with $t$.
The roots $e_1$ and $e_2$ acquire an imaginary part for $m^2 < t < 9m^2$.
At $t=0$ the roots $e_2$ and $e_3$ coincide.
At $t=m^2$ and at $t=9m^2$ the roots $e_1$ and $e_2$ coincide.
At $t=\infty$ the roots $e_1$ and $e_3$ coincide.
In the interval $m^2 \le t \le 9 m^2$ the real parts of the roots $e_1$ and $e_2$ coincide.
}
\end{figure}
This is shown in fig.~\ref{fig_roots}.
At the singular points $t \in \{0,m^2,9m^2,\infty\}$  two of the three roots coincide.
In detail we have
\begin{center}
\begin{tabular}{|c|cccc|}
\hline
 $t$    & $0$       & $m^2$  & $9m^2$  & $\infty$ \\
\hline
 roots  & $e_2=e_3$ & $e_1=e_2$ & $e_1=e_2$ & $e_1=e_3$ \\
\hline
\end{tabular}
\end{center}
The modulus $k$ and the complementary modulus $k'$ of the elliptic curve are defined by
\bq
\label{def_modulus}
 k^2 = \frac{e_3-e_2}{e_1-e_2},
 \qquad
 k'{}^2 = 1-k^2 = \frac{e_1-e_3}{e_1-e_2}.
\eq
The modulus $k$ and the complementary modulus $k'$ appear as arguments of
the complete elliptic integral of the first kind $K(k)$, defined by
\bq
\label{def_K}
 K(k)
 & = &
 \int\limits_0^1 \frac{dt}{\sqrt{\left(1-t^2\right)}\sqrt{\left(1-k^2t^2\right)}}.
\eq
It is clear from the definition that $K(k)$ is only a function of $k^2$ and some authors prefer therefore the notation $K(k^2)$.
In this paper we follow the standard conventions \cite{NIST:Handbook} and define $K(k)$ as in eq.~(\ref{def_K}).
The complete elliptic integral of the first kind $K(k)$ has branch cuts on the real axis at $]-\infty,-1]$ and $[1,\infty[$
in the complex $k$-plane.
The function $\tilde{K}(k^2)=K(k)$ has then a branch cut at $[1,\infty[$ in the complex $k^2$-plane.

Let us denote by $\delta_1$ and $\delta_2$ two cycles on the elliptic curve $E_t$ 
which generate the homology group $H_1(E_t,{\mathbb Z})$.
The differential forms
\bq
 \eta_1 \; = \; \frac{dx}{y},
 & &
 \eta_2 \; = \; \frac{x dx}{y}
\eq
are generators of the cohomology group $H^1_{\mathrm{dR}}(E_t)$.
The periods of the elliptic curve are
\bq
 P_{ij} & = & \int\limits_{\delta_i} \eta_j,
 \;\;\;\;\;\;\;\;\; i,j \in \{1,2\}.
\eq
We are in particular interested in the periods $\psi_1=P_{11}$ and $\psi_2=P_{21}$, involving $\eta_1$.
These periods satisfy the differential equation
\bq
\label{diff_eq_periods}
 \left[ 
    \frac{d^2}{dt^2} 
    + \left( \frac{1}{t} + \frac{1}{t-m^2} + \frac{1}{t-9m^2} \right) \frac{d}{dt} 
    + \frac{1}{m^2} \left( - \frac{1}{3 t} + \frac{1}{4\left(t-m^2\right)} + \frac{1}{12\left(t-9m^2\right)} \right)
 \right] \psi_{i} & = & 0.
 \nonumber \\
\eq
(The periods $P_{12}$ and $P_{22}$ satisfy a slightly different differential equation.)
Eq.~(\ref{diff_eq_periods}) is called the Picard-Fuchs equation.
We define the two cycles $\delta_1$ and $\delta_2$ such that the periods $\psi_1$ and $\psi_2$
are given for $t<0$ by
\bq
\label{def_periods_I}
 \psi_1 =  
 2 \int\limits_{e_2}^{e_3} \frac{dx}{y},
 & &
 \psi_2 =  
 2 \int\limits_{e_1}^{e_3} \frac{dx}{y}.
\eq
We take
\bq
 y & = & - 2 \sqrt{x-e_1} \sqrt{x-e_2} \sqrt{x-e_3}.
\eq
The path of integration in $x$-space is such that we have an infinitesimal small negative imaginary part for $x$.
In other words, the integration path is below all cuts.
Working out the integrals one finds in the region $t<0$
\bq
\label{def_periods_II}
 \psi_1 
  =
 \frac{4 \mu^2}{\left( m^2 - t\right)^{\frac{3}{4}} \left( 9 m^2 - t \right)^{\frac{1}{4}}} K\left(k\right),
 & &
 \psi_2 =  
 \frac{4 i \mu^2}{\left( m^2 - t\right)^{\frac{3}{4}} \left( 9 m^2 - t \right)^{\frac{1}{4}}} K\left(k'\right).
\eq
The periods $\psi_1$ and $\psi_2$ vary continuously with $t$.
\begin{figure}
\begin{center}
\includegraphics[scale=1.0]{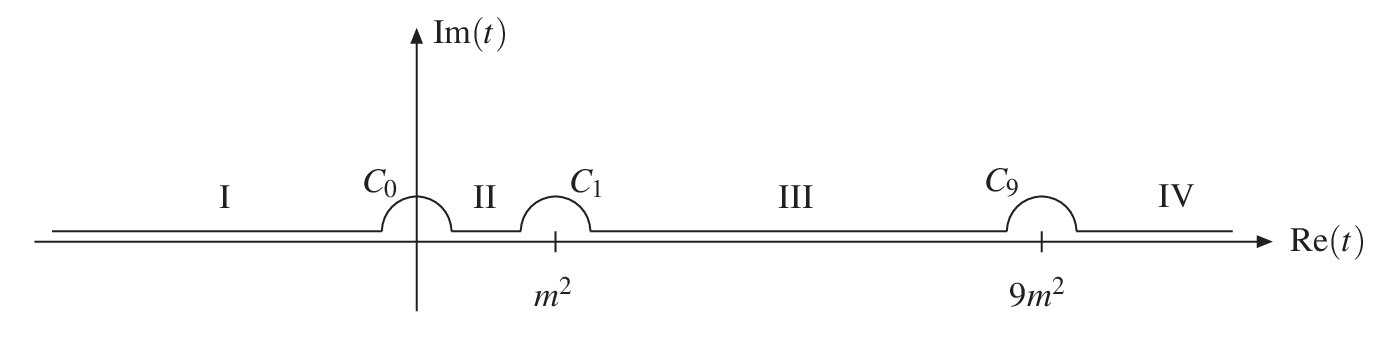}
\end{center}
\caption{\label{fig_t_path}
The path for the analytic continuation in the variable $t$. Feynman's $i0$-prescription avoids the singular points 
at $0$, $m^2$ and $9m^2$ as shown in the figure.
}
\end{figure}
We study the variation of these periods along the path shown in fig.~(\ref{fig_t_path}).
We divide the path into seven pieces: Four line segments and three small semi-circles.
The line segments are characterised by
\bq
\begin{array}{lrcccr}
 \mbox{Region I}: & -\infty & < & \mathrm{Re}(t) & < & 0, \\
 \mbox{Region II}: &       0 & < & \mathrm{Re}(t) & < & m^2, \\
 \mbox{Region III}: &     m^2 & < & \mathrm{Re}(t) & < & 9 m^2, \\
 \mbox{Region IV}: &   9 m^2 & < & \mathrm{Re}(t) & < & \infty. \\
 \end{array}
\eq
The three semi-circles $C_0$, $C_1$ and $C_9$ encircle the points $0$, $m^2$ and $9m^2$, respectively.
Everywhere on the path we have $\mathrm{Im}(t) > 0$. This implements Feynman's $i0$-prescription.
Of course, the path is equivalent to a straight line parallel to the real axis with a small imaginary part.
However, it is advantageous to discuss the individual pieces separately, in particular the small semi-circles.
We may extend the path to a (closed) path on the Riemann sphere by adding a semi-circle at infinity.
As already mentioned, the periods $\psi_1$ and $\psi_2$ are continuous functions of $t$.
In eq.~(\ref{def_periods_II}) we expressed the two periods for $t$ in region I in terms of complete
elliptic integrals of the first kind. The complete elliptic integral $K(k)$ has branch cuts on the real axis.
In terms of the variable $k^2$ the branch cut is on the positive real axis given by the interval $[1,\infty[$.
In fig.~(\ref{fig_k2_path}) we sketch
\begin{figure}
\begin{center}
\includegraphics[scale=1.0]{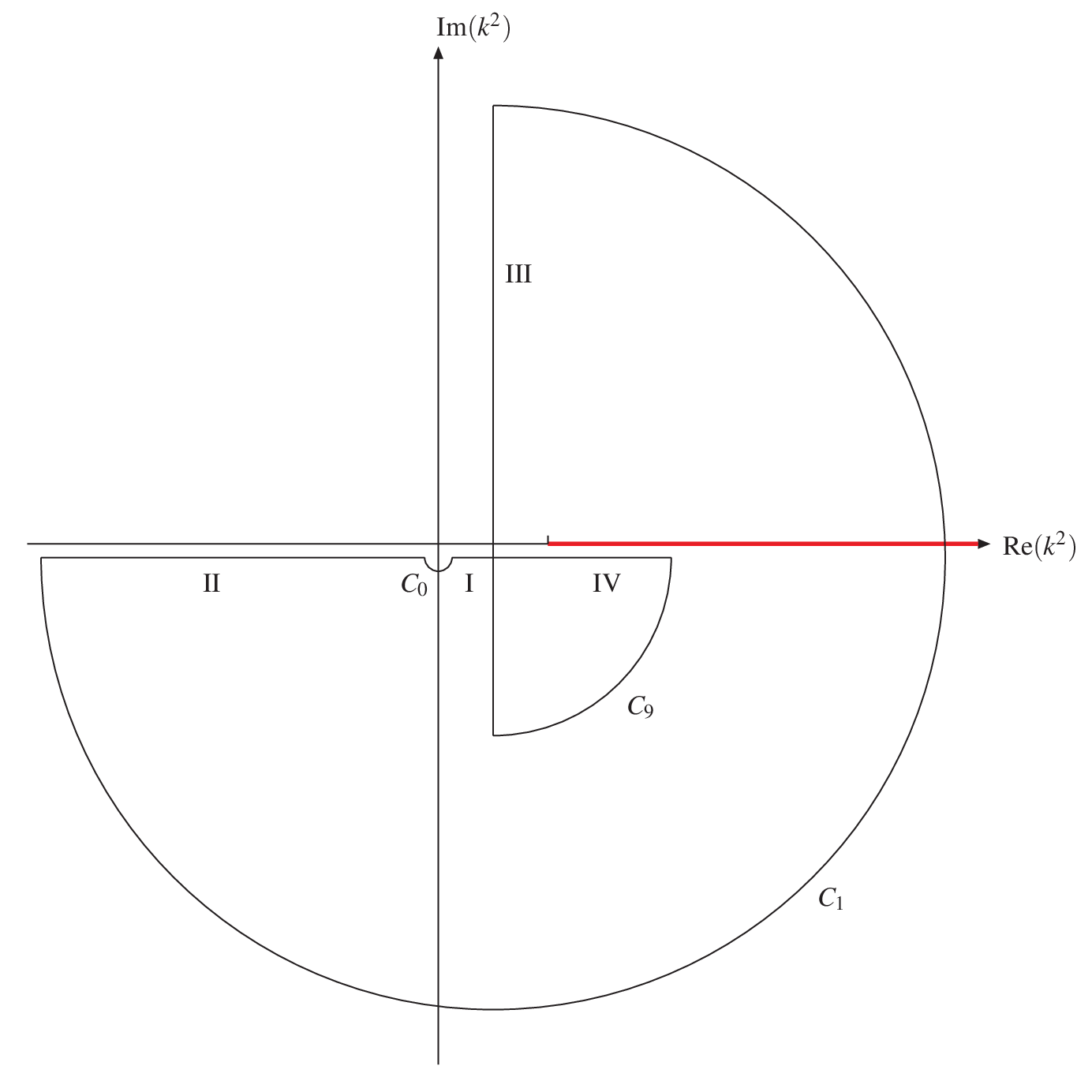}
\end{center}
\caption{\label{fig_k2_path}
The path in the complex $k^2$-space, as $t$ varies along the path of fig.~(\ref{fig_t_path}).
In the regions IV, I and II the path in $k^2$-space is at an infinitesimal distance below the real axis.
The path crosses the branch cut $[1,\infty]$ in the region $C_1$.
In region III we have $\mathrm{Re}(k^2)=1/2$.
}
\end{figure}
the path in the complex $k^2$-space, as $t$ varies along the path 
shown in fig.~(\ref{fig_t_path}).
For $t=-\infty+i0$ we start in $k^2$-space at $k^2=1-i0$, continuing below the real axis until we reach for
$t=m^2+i0$ the point $k^2=-\infty-i0$.
The small semi-circle $C_0$ around $t=0$ is harmless and can be deformed away.
However, the semi-circle $C_1$ around $t=m^2$ is essential. 
It corresponds to a three-quarter circle in $k^2$-space and crosses
the branch cut $[1,\infty[$.
The path continues in region III, where we have $\mathrm{Re}(k^2)=1/2$.
The semi-circle around $t=9m^2$ brings the path close to the real axis with $\mathrm{Re}(k^2)>1$.
In region IV the path continues at an infinitesimal distance below the real axis back to $k^2=1-i0$.
The three-quarter circle in fig.~(\ref{fig_k2_path}) corresponding to $C_1$ is equivalent
to the missing quarter-circle in the clockwise direction and a full circle in the anti-clockwise direction.
In this way the monodromy of $K(k)$ enters the expression for $\psi_1$.

\begin{figure}
\begin{center}
\includegraphics[scale=1.0]{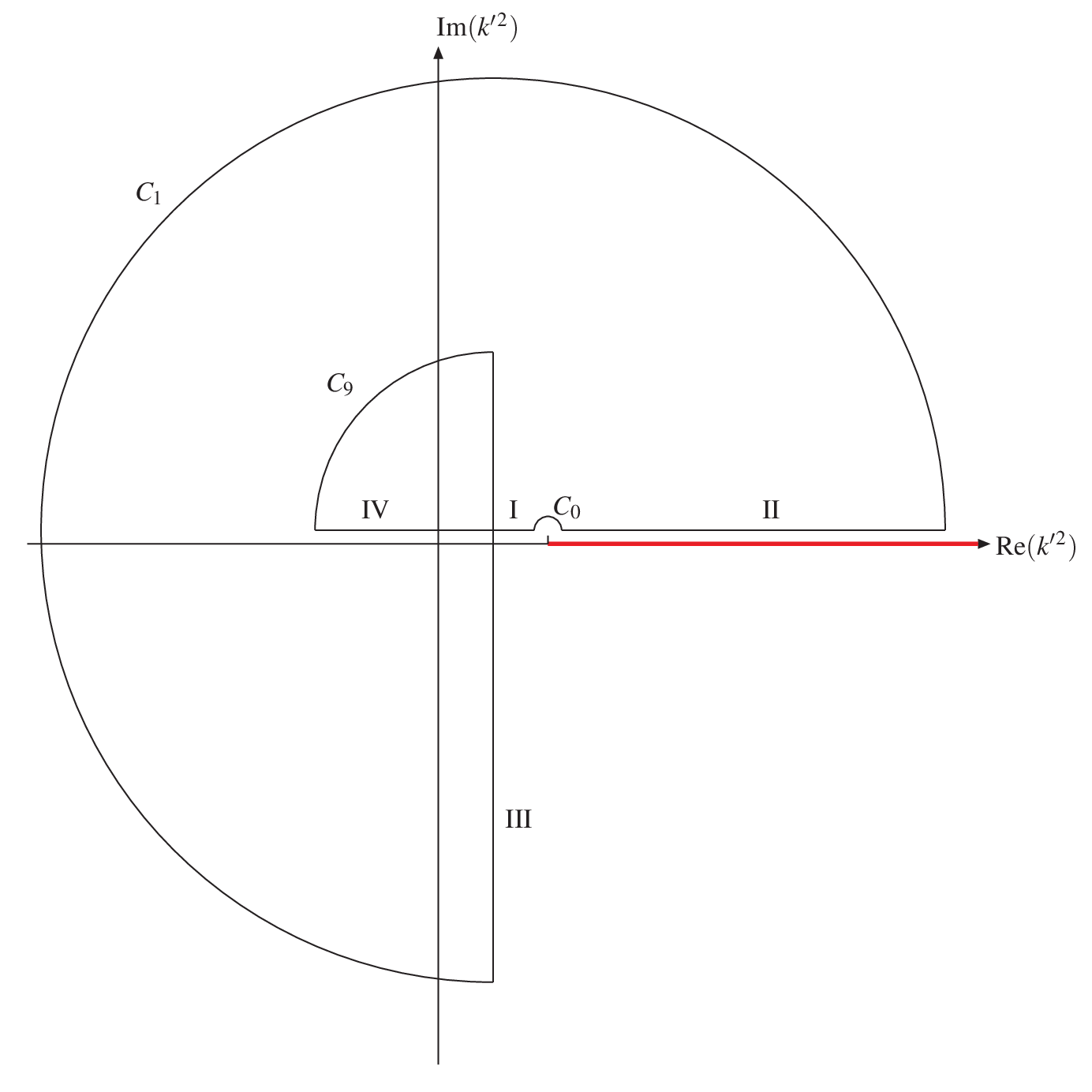}
\end{center}
\caption{\label{fig_kprime2_path}
The path in the complex $k'{}^2$-space, as $t$ varies along the path of fig.~(\ref{fig_t_path}).
In the regions IV, I and II the path in $k'{}^2$-space is at an infinitesimal distance above the real axis.
In region III we have $\mathrm{Re}(k^2)=1/2$.
}
\end{figure}

For $\psi_2$ we study the path in $k'{}^2$-space, as $t$ varies along the path of fig.~(\ref{fig_t_path}).
Due to the relation
\bq
 k^2 + k'{}^2 & = & 1,
\eq
the path in $k'{}^2$-space is simply obtained from the path in $k^2$-space by reflection on the point $1/2$.
The path in $k'{}^2$-space is sketched in fig.~(\ref{fig_kprime2_path}).
Let us stress that the path in $k'{}^2$-space does not cross the branch cut $[1,\infty]$.

\begin{figure}
\begin{center}
\includegraphics[scale=1.0]{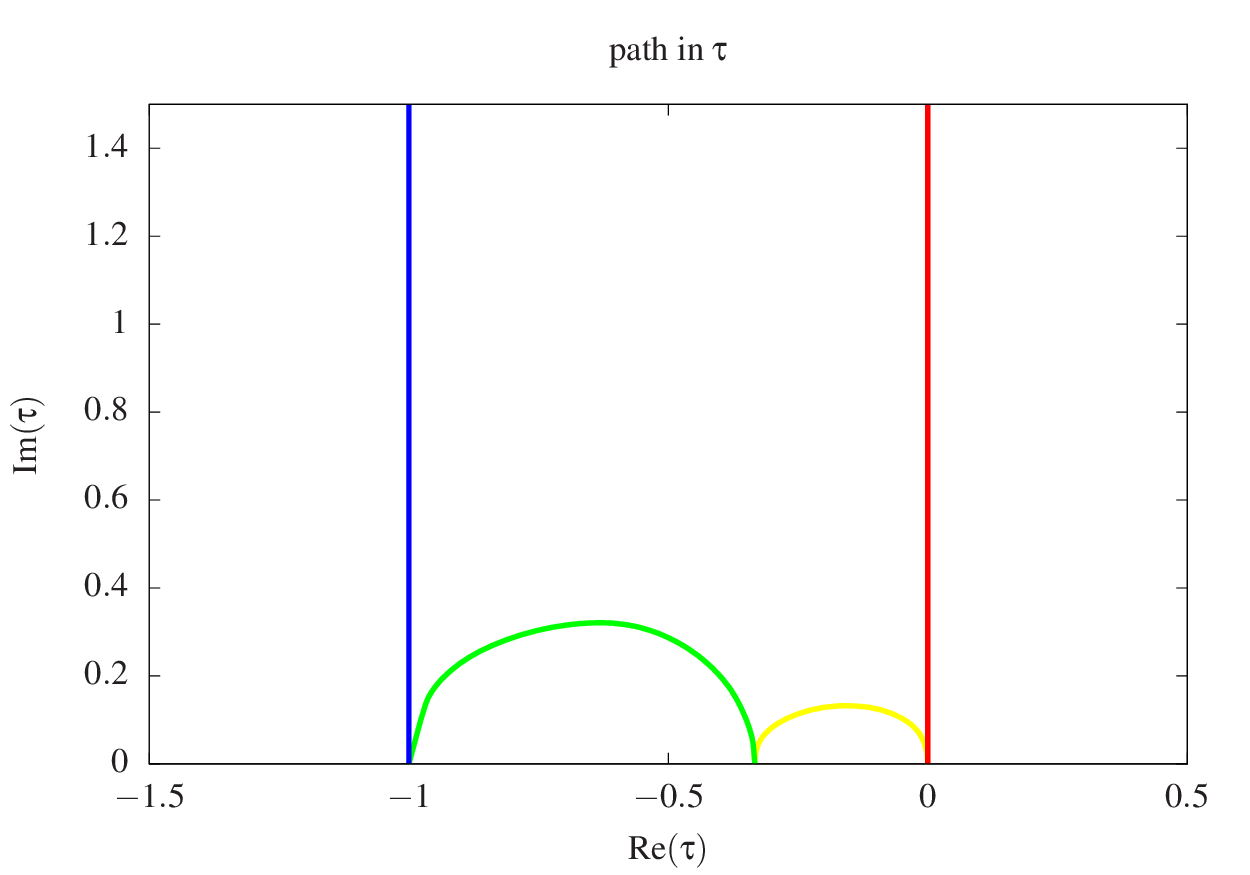}
\includegraphics[scale=1.0]{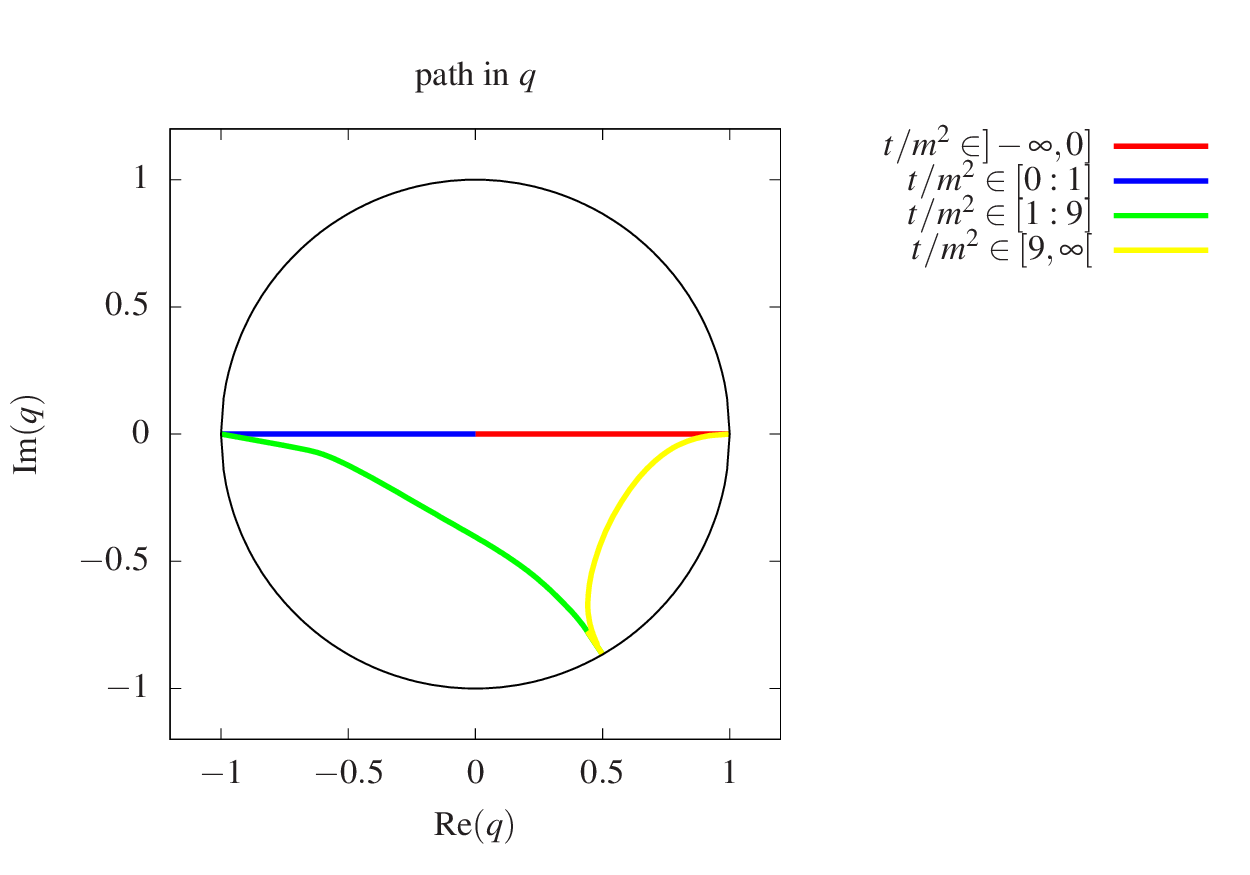}
\end{center}
\caption{\label{fig_tau_nome}
The path in $\tau$-space and in $q$-space, as $t$ varies along the path of fig.~(\ref{fig_t_path}).
The value $t=-\infty$ corresponds to $\tau=0$ and $q=1$,
the value $t=0$ corresponds to $\tau=i\infty$ and $q=0$,
the value $t=m^2$ corresponds to $\tau=-1$ and $q=-1$,
the value $t=9m^2$ corresponds to $\tau=-\frac{1}{3}$ and $q=\frac{1}{2} - \frac{i \sqrt{3}}{2}$.
}
\end{figure}

In expressing the periods $\psi_1$ and $\psi_2$ for all values of $t$ along the path of fig.~(\ref{fig_t_path})
in terms of complete elliptic integrals of the first kind $K(k)$ and $K(k')$ we have to take the monodromy at
$t=m^2$ into account.
The monodromy relation is derived in section~\ref{sect:picard_lefschetz}.
The result is as follows:
For all values of $t \in {\mathbb R}$ the periods $\psi_1$ and $\psi_2$
are expressed as
\bq
\label{def_periods_final}
 \left( \begin{array}{c}
 \psi_2\left(t+i0\right) \\
 \psi_1\left(t+i0\right) \\
 \end{array} \right)
 & = &
 \frac{4 \mu^2}{\left( m^2 - t - i0 \right)^{\frac{3}{4}} \left( 9 m^2 - t - i0 \right)^{\frac{1}{4}}}
 \; \gamma_t \;
   \left( \begin{array}{c}
     i K\left(k'\left(t+i0\right)\right) \\
     K\left(k\left(t+i0\right)\right) \\
   \end{array} \right),
\eq
where the $2\times2$ matrix $\gamma_t$ is given by
\bq
\label{monodromy}
 \gamma_t
 & = &
 \left\{
  \begin{array}{rc}
 \left( \begin{array}{rr}
  1 & 0 \\
  0 & 1 \\
 \end{array} \right),
   &
   -\infty < t < m^2, \\
 & \\
 \left( \begin{array}{rr}
  1 & 0 \\
  -2 & 1 \\
 \end{array} \right),
   &
   m^2 < t < \infty, \\
  \end{array}
 \right.
\eq
and encodes the monodromy.
Note that in region II the elliptic integral $K(k')$ is evaluated above the branch cut,
while in region IV the elliptic integral $K(k)$ is evaluated below the branch cut.
In both cases we may use for $k \in ]1,\infty[$ the formula \cite{NIST:Handbook}
\bq
 K\left(k \pm i 0 \right)
 & = &
 \frac{1}{k}
  \left[ 
        K\left(\frac{1}{k}\right) \pm i K\left(\sqrt{1-\frac{1}{k^2}}\right)
  \right].
\eq
We note that this relation can also be used for an analytical continuation
by fixing the constants in a small neighbourhood around the singular points \cite{Remiddi:2016gno,Primo:2017ipr}.
However, we will pursue a different (and simpler) path here. 
Having defined the periods $\psi_1$ and $\psi_2$ in eq.~(\ref{def_periods_final}),
we introduce the standard definitions of the ratio of the two periods $\tau$ and the nome $q$ of the elliptic curve:
\bq
 \tau 
 \;\; = \;\;
 \frac{\psi_2}{\psi_1},
 \;\;\;
 & &
 \;\;\;
 q 
 \;\; = \;\;
 e^{i \pi \tau}.
\eq
In fig.~(\ref{fig_tau_nome}) we show the path in $\tau$-space and in $q$-space
as $t$ varies along the path of fig.~(\ref{fig_t_path}).
We have
\bq
\label{def_tau_nome}
 \tau(t=\pm\infty) = 0,
 \;\;\;
 \tau(t=0) = i \infty,
 \;\;\;
 \tau(t=m^2) = -1,
 \;\;\;
 \tau(t=9m^2) = - \frac{1}{3}.
\eq
For all values $t$ along the path of fig.~(\ref{fig_t_path}) we have
\bq
\label{upper_half}
 \mathrm{Im}(\tau) > 0,
 & &
 \left| q \right| < 1.
\eq
We have
\bq
 \mathrm{Im}(\tau) = 0,
 & &
 \left| q \right| = 1.
\eq
for $t \in \{m^2, 9m^2, \pm\infty \}$, i.e. without Feynman's $i0$-prescription.
There is a bijection between all points $t$ on the path of fig.~(\ref{fig_t_path})
and the points $\tau$ on the path in $\tau$-space of fig.~(\ref{fig_tau_nome}).
In the direction $t \rightarrow \tau$ the mapping is given by eq.~(\ref{def_tau_nome})
\bq
\label{mapping_t_to_tau}
 \tau & = & \frac{\psi_2(t)}{\psi_1(t)}.
\eq
In the reverse direction we have
\bq
\label{mapping_tau_to_t}
 t & = &
 - 9 m^2 
 \frac{\eta\left(\tau\right)^4 \eta\left(\frac{3\tau}{2}\right)^4 \eta\left(6\tau\right)^4}
      {\eta\left(\frac{\tau}{2}\right)^4 \eta\left(2\tau\right)^4 \eta\left(3\tau\right)^4},
\eq
where $\eta(\tau)$ denotes Dedekind's eta function, defined by
\bq
 \eta (\tau) & = & e^{\frac{i \pi \tau}{12}} \prod\limits_{n=1}^{\infty} (1-e^{2\pi i n \tau}).
\eq
We have checked numerically that eq.~(\ref{mapping_tau_to_t}) is the inverse mapping to eq.~(\ref{mapping_t_to_tau})
for all $t$ on the path of fig.~(\ref{fig_t_path}).

\section{Analytic continuation of Feynman integrals}
\label{sec:feynman_integrals}

In this section we discuss the analytic continuation of the Feynman integrals associated to the kite
family, i.e. the kite integral and all sub-topologies, including in particular
the equal mass sunrise integral.
We will see that this is trivial, once the periods $\psi_1$ and $\psi_2$ are defined by
eq.~(\ref{def_periods_final}).
The Feynman integrals of the kite family depend on the variable 
\bq
 t & = & p^2.
\eq
Feynman's $i0$-prescription instructs us to add a small imaginary part $t \rightarrow t + i0$ where necessary.
We are interested in the values of these integrals as $t$ ranges over the real numbers.
Using integration-by-parts identities \cite{Tkachov:1981wb,Chetyrkin:1981qh}
we may express all integrals from this family as linear
combinations of a few master integrals.
For the kite family there are eight master integrals \cite{Remiddi:2016gno,Adams:2016xah}, which we denote
as $\vec{I}=(I_1,I_2,...,I_8)$.
We will follow the notation of \cite{Adams:2016xah}, where the definition of $I_1$-$I_8$ is given.
The eight master integrals $\vec{I}$ for the kite system satisfy a system of differential equations
in $t$ of Fuchsian type \cite{Kotikov:1990kg,Kotikov:1991pm,Remiddi:1997ny,Gehrmann:1999as,Argeri:2007up,MullerStach:2012mp,Henn:2013pwa,Henn:2014qga,Adams:2017tga}, which has singularities at
\bq
 \left\{ 0, m^2, 9 m^2, \infty \right\}.
\eq
The system of differential equations reads
\bq
 \label{diff_eq_kite_1}
 \mu^2 \frac{d}{dt} \vec{I}
 & = &
 \left[ 
  \frac{\mu^2}{t} A_0 + \frac{\mu^2}{t-m^2} A_1 + \frac{\mu^2}{t-9m^2} A_9 \right] \vec{I},
\eq
where $A_0$, $A_1$ and $A_9$ are $8 \times 8$-matrices with entries of the form $a+b\eps$, where $a,b \in {\mathbb Q}$
and $\eps$ denotes the dimensional regularisation parameter.
The explicit expressions are given in \cite{Adams:2016xah}.
The system of differential equations in eq.~(\ref{diff_eq_kite_1}) holds for all $t \in {\mathbb C}$.
In \cite{Adams:2016xah} we solved the system of differential equations in the Euclidean region (i.e. region I) 
by performing a change of variables from $t$ to the nome $q$.
For this change of variables the relation between $t$ and $\tau$ is given by
\bq
\label{relation_t_tau}
 t = f(\tau),
 & &
 f(\tau) =
 - 9 m^2 
 \frac{\eta\left(\tau\right)^4 \eta\left(\frac{3\tau}{2}\right)^4 \eta\left(6\tau\right)^4}
      {\eta\left(\frac{\tau}{2}\right)^4 \eta\left(2\tau\right)^4 \eta\left(3\tau\right)^4}.
\eq
The relation between $\tau$ and $q$ is as usual $q=\exp(i \pi \tau)$.
In terms of the variable $q$ the system of differential equations~(\ref{diff_eq_kite_2}) becomes
\bq
 \label{diff_eq_kite_2}
 q \frac{d}{dq} \vec{I}
 & = &
 \left(
  g_{2,0} A_0
  +
  g_{2,1} A_1
  +
  g_{2,9} A_9
 \right) \vec{I},
\eq
where $g_{2,0}$, $g_{2,1}$ and $g_{2,9}$ are modular forms of modular weight $2$ for the congruence subgroup $\Gamma_0(12)$,
defined in \cite{Adams:2017ejb}.

We have seen in eq.~(\ref{mapping_tau_to_t}) that the validity of eq.~(\ref{relation_t_tau})
is not restricted to region I, but holds for all values $t$ on the path of fig.~(\ref{fig_t_path}).
Thus eq.~(\ref{diff_eq_kite_2}) holds for all values of $q$ on the path
of fig.~(\ref{fig_tau_nome}).
We already know a solution for this system of differential equations: In \cite{Adams:2016xah} we solved these differential
equations in terms of $\mathrm{ELi}$-functions in the region I.
The solution extends to all values $t$ on the path of fig.~(\ref{fig_t_path}).
Furthermore from eq.~(\ref{upper_half}) it follows that
\bq
 \left| q \right| \;\; < \;\; 1
 & & \mbox{for} \; t \in {\mathbb R} \backslash \{m^2,9m^2,\infty\}.
\eq
This ensures the convergence of the $q$-series, which we use to express our results.

Let us give an example: The first term of the Laurent expansion in the dimensional regularisation parameter $\eps$
of the equal-mass sunrise integral in two space-time dimensions reads
\bq
\label{example_sunrise}
 S_{111}^{(0)}(2,t)
 & = &
 \frac{3 \psi_1}{i \pi}
 \left\{
 \frac{1}{2} \mathrm{Li}_2\left( r_3 \right) - \frac{1}{2} \mathrm{Li}_2\left( r_3^{-1} \right)
 + \mathrm{ELi}_{2;0}\left(r_3;-1;-q\right)
 - \mathrm{ELi}_{2;0}\left(r_3^{-1};-1;-q\right)
 \right\},
 \nonumber \\
\eq
where $r_3=\exp(2\pi i/3)$ and the functions $\mathrm{ELi}_{n;m}(x,y,q)$ are defined by
\bq
 \mathrm{ELi}_{n;m}\left(x;y;q\right) & = & 
 \sum\limits_{j=1}^\infty \sum\limits_{k=1}^\infty \; \frac{x^j}{j^n} \frac{y^k}{k^m} q^{j k}.
\eq
Eq.~(\ref{example_sunrise}) holds for all values of $t$, the periods $\psi_1$ and $\psi_2$ (and in turn the nome $q$) 
are computed by eq.~(\ref{def_periods_final}).

Let us summarise: With the definition of the periods $\psi_1$ and $\psi_2$ 
given in eq.~(\ref{def_periods_final}) the differential equation~(\ref{diff_eq_kite_2}) holds for all
values of $t \in {\mathbb R} + i0$.
Thus any solution $\vec{I}(t)$, originally defined in one particular region (say region I),
extends to all values $t \in {\mathbb R} + i0$.
Furthermore we have for $t \in {\mathbb R} + i0$
\bq
 \left| q \right| & < & 1,
\eq
ensuring the convergence of the $\mathrm{ELi}$-series.
(At the singular points $t \in \{m^2, 9 m^2, \infty \}$ we have $|q|=1$.)
In other words: The analytic continuation of $\vec{I}(t)$ could not be simpler!


\section{Monodromy and the Picard-Lefschetz theorem}
\label{sect:picard_lefschetz}

\begin{figure}
\begin{center}
\includegraphics[scale=0.55]{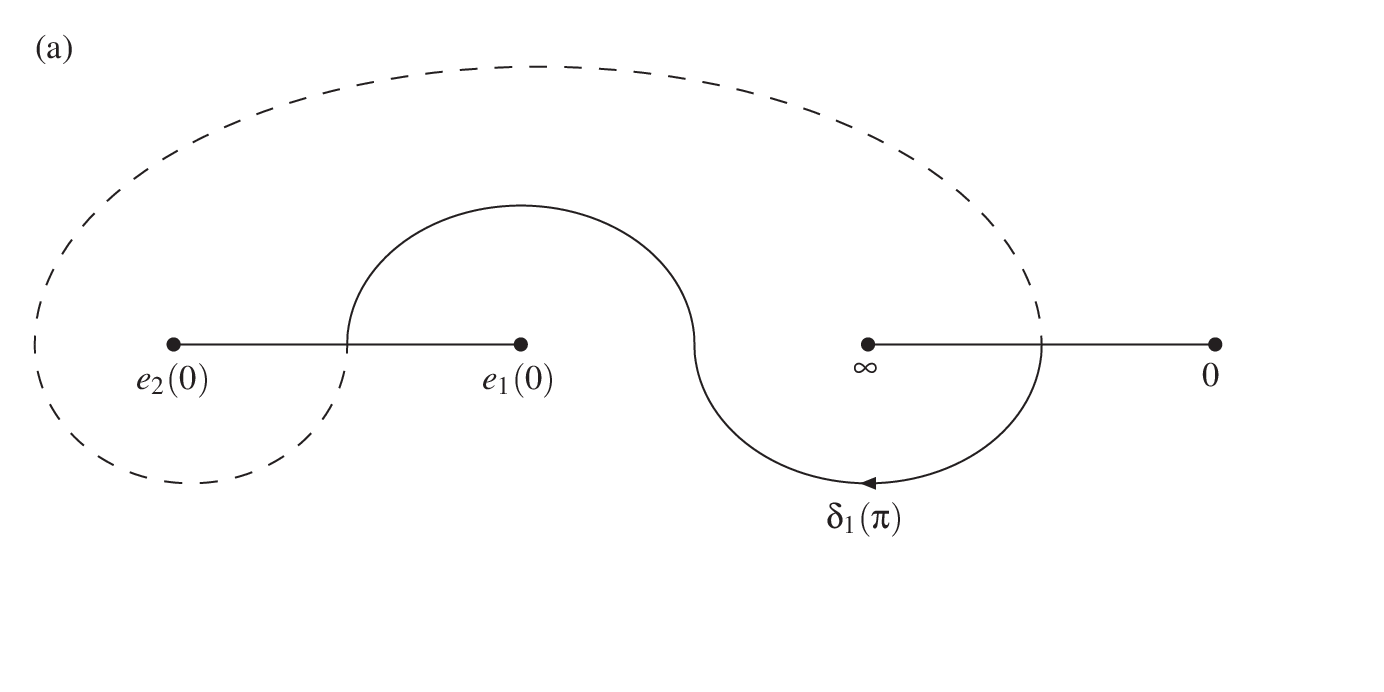}
\hspace*{5mm}
\includegraphics[scale=0.55]{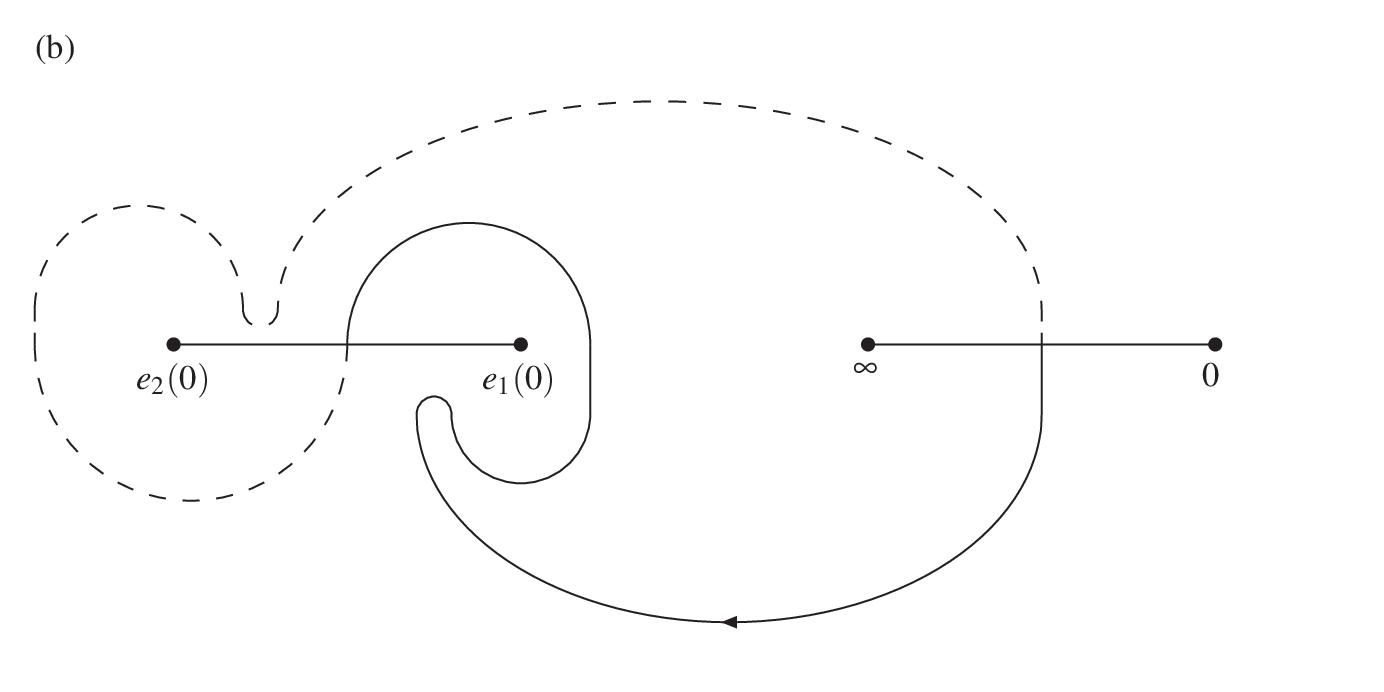}
\includegraphics[scale=0.55]{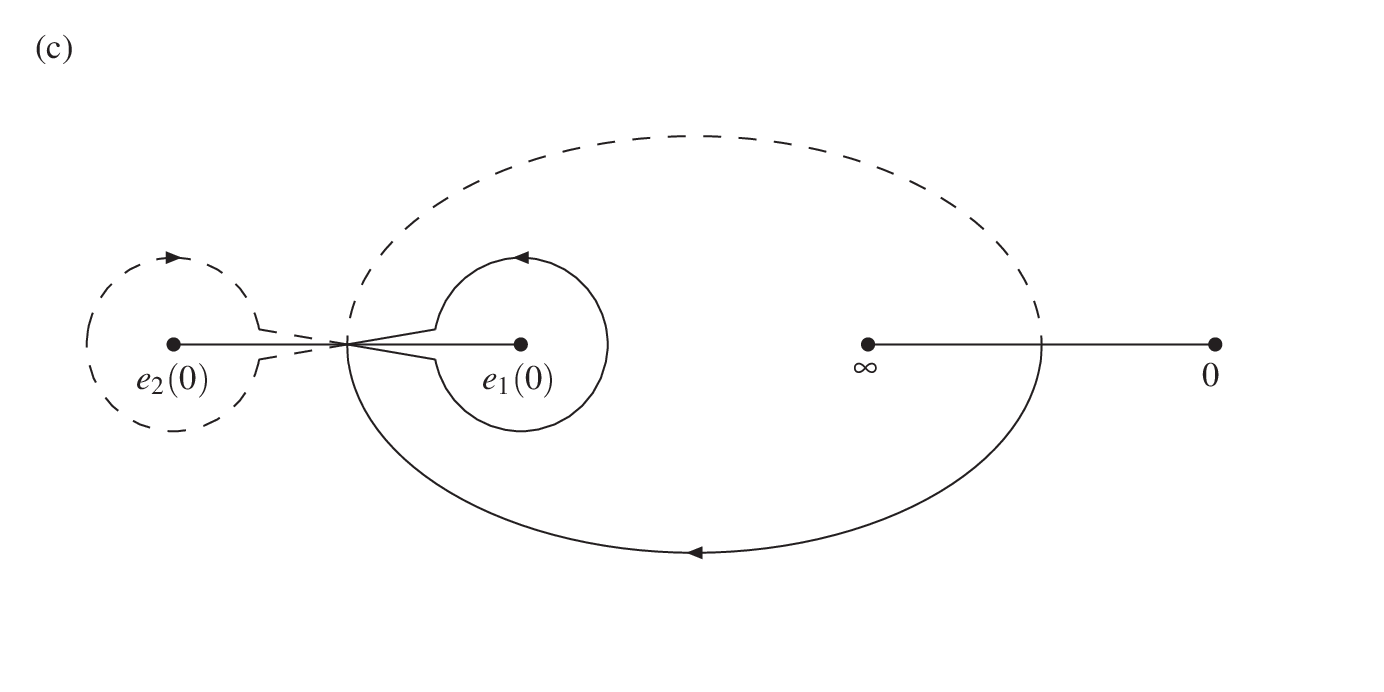}
\hspace*{5mm}
\includegraphics[scale=0.55]{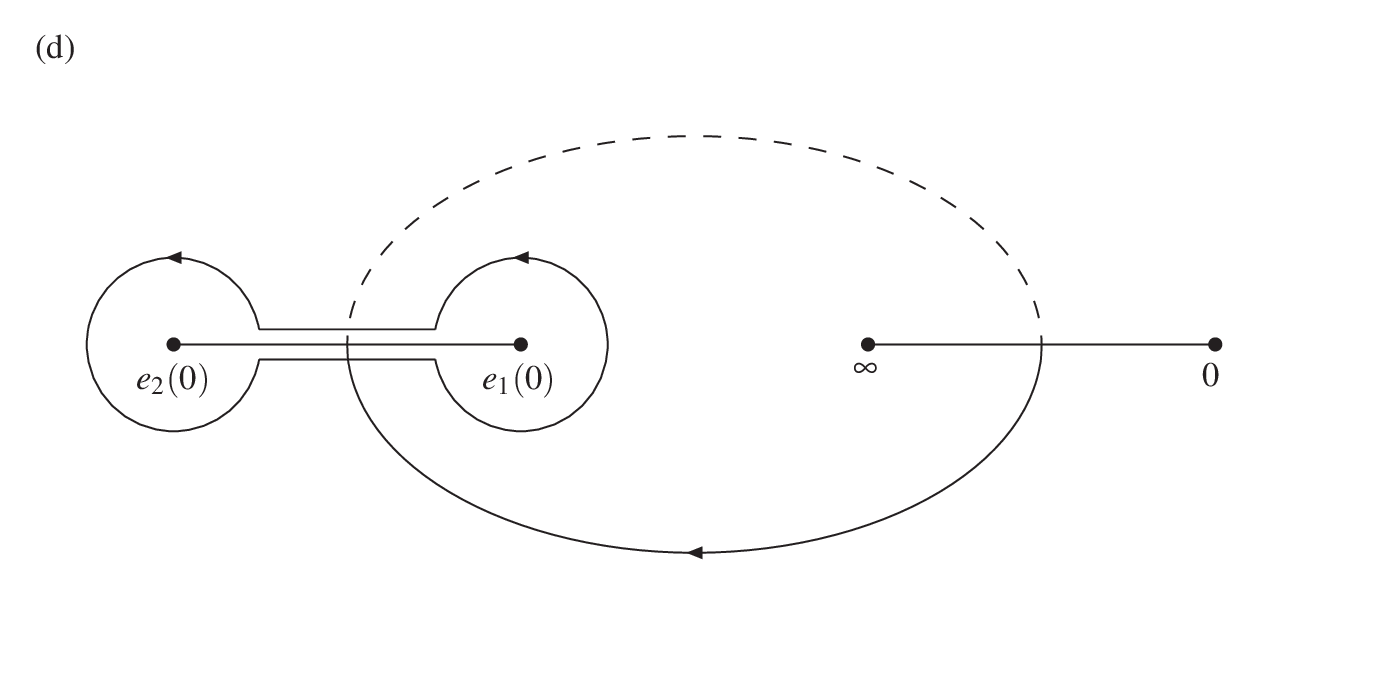}
\includegraphics[scale=0.55]{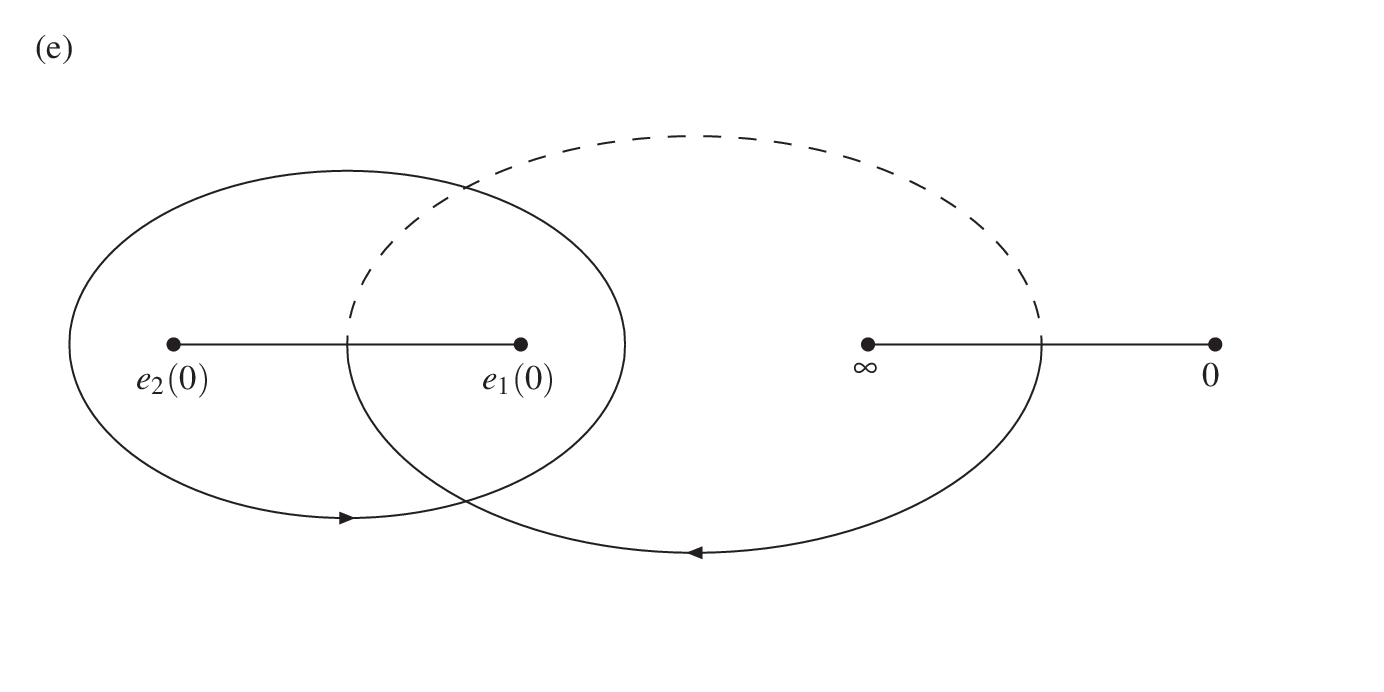}
\end{center}
\caption{\label{fig_deformation_delta1}
Deformation of the cycle $\delta_1(\pi)$.
In the step from $(c)$ to $(d)$ we bring the closed path around $e_2(0)$ from one Riemann sheet to the other
Riemann sheet. This changes the orientation. The last figure shows that $\delta_1(\pi)=\delta_1(0)-\delta_2(0)$.
}
\end{figure}

In this section we derive the monodromy relation of eq.~(\ref{def_periods_final}) and eq.~(\ref{monodromy}).
This is a standard application of the Picard-Lefschetz theorem and we follow the textbooks \cite{Carlson,Hwa}.
We may deform the path $C_1$ in fig.~(\ref{fig_k2_path}) in $k^2$-space into 
a quarter-circle in the clockwise direction and a small full circle around $1$ in the anti-clockwise direction,
\begin{figure}
\begin{center}
\includegraphics[scale=0.35]{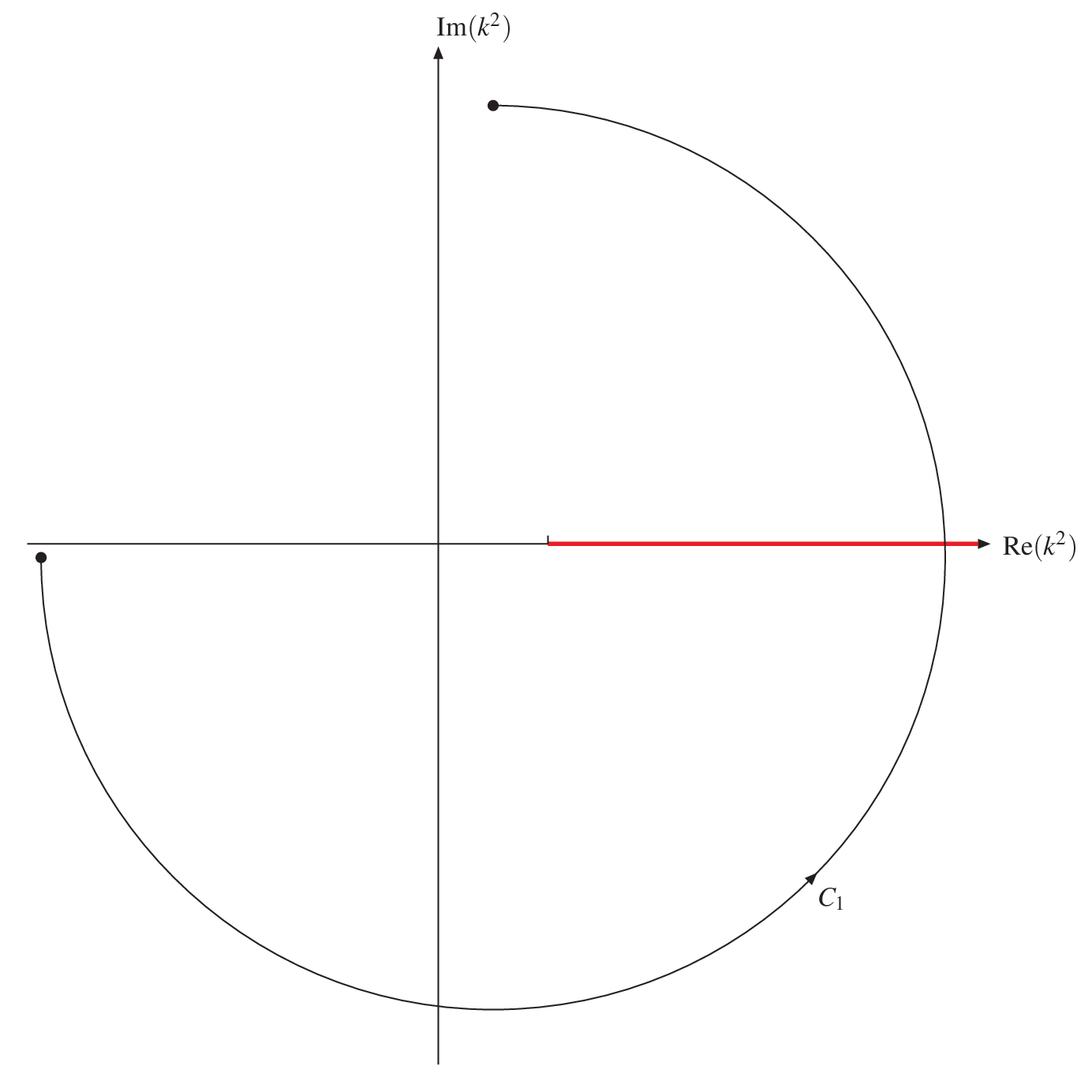}
\hspace*{5mm}
\includegraphics[scale=0.35]{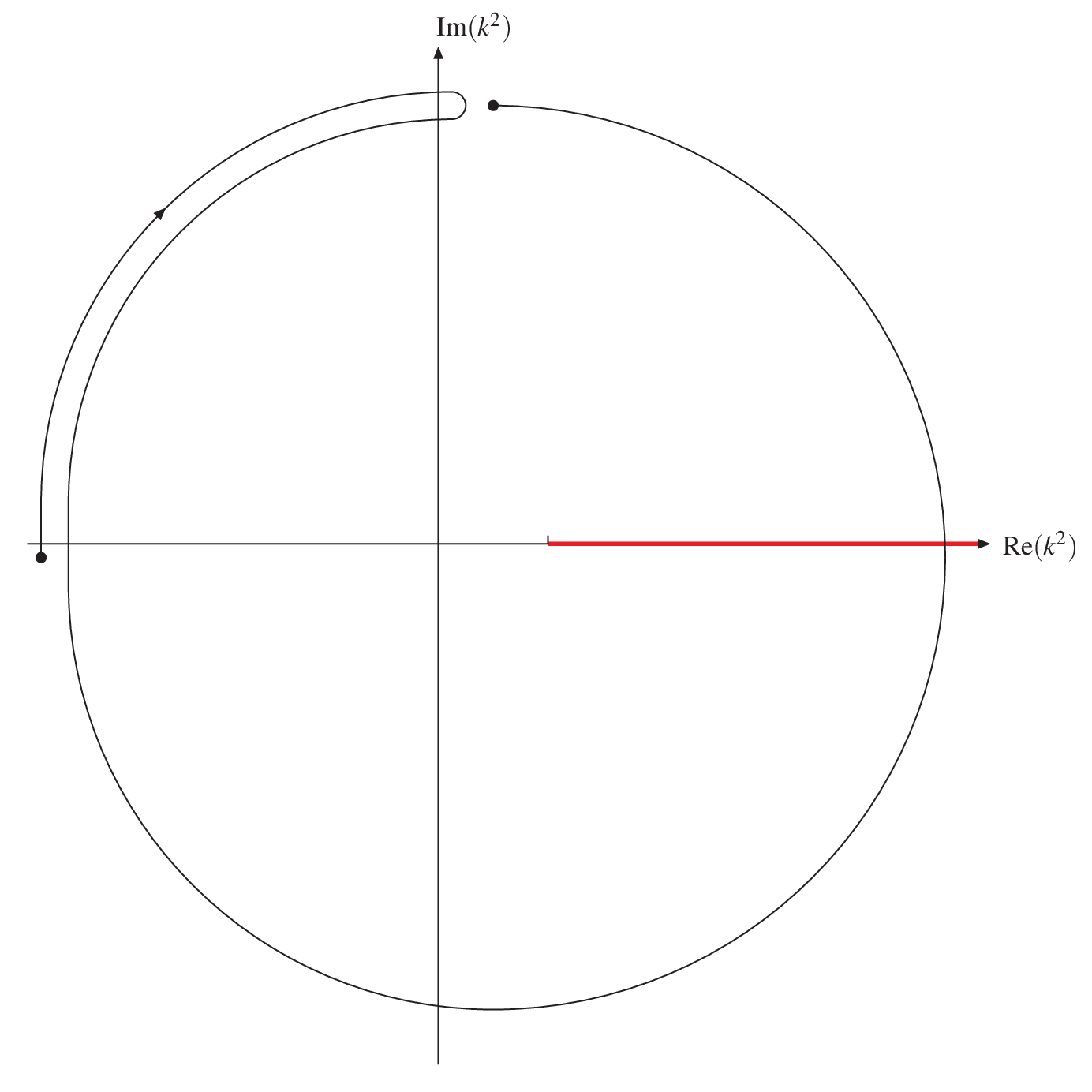}
\hspace*{5mm}
\includegraphics[scale=0.35]{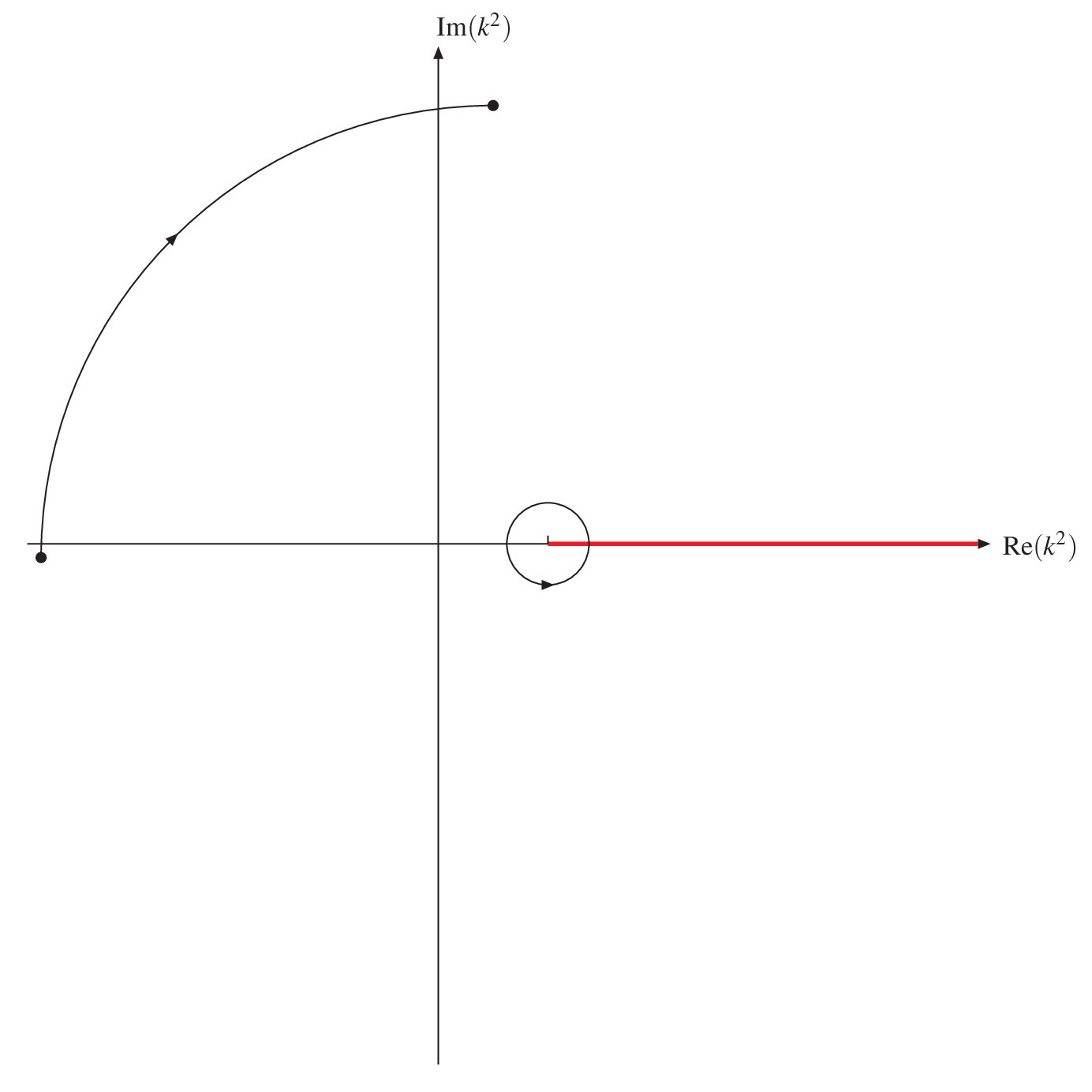}
\end{center}
\caption{\label{fig_k2_deform}
Deformation of the path $C_1$ in $k^2$-space into a quarter-circle in the clockwise direction and a small full circle
around $1$ in the counter-clockwise direction.
}
\end{figure}
as shown in fig.~\ref{fig_k2_deform}.
For the monodromy we have to study the contribution from the small full circle around $1$.
Let us set $\lambda=k^2$.
We recall that we may write the complete elliptic integral of the first kind equally well as
\bq
 K(k)
 & = &
 \frac{1}{2} \int\limits_0^{\lambda} \frac{dt}{\sqrt{t} \sqrt{\lambda-t} \sqrt{1-t}}.
\eq
Thus, $K(k)$ gives a quarter-period of the elliptic curve in Legendre form
\bq
\label{Legendre_form}
 E_\lambda & : &
 y^2 \; = \; p(x),
 \;\;\;
 p(x) = x \left(x-\lambda\right) \left(x-1\right).
\eq
The roots of the polynomial $p(x)$ are 
\bq
 0, \;\; \lambda, \;\; 1, \;\; \infty.
\eq
Let us now study the behaviour of $K(k)$ as $\lambda$ moves in a small circle around $1$.
It is slightly simpler to consider instead of the family of elliptic curves $E_\lambda$
the family of elliptic curves $E_\varphi$ given by
\bq
 E_\varphi & : & 
 y^2 \; = \;  p(x,\varphi),
 \;\;\;
 p(x,\varphi) \; = \; x \left(x-e_1(\varphi)\right) \left(x-e_2(\varphi)\right),
\eq
where the roots are given by
\bq
 e_1(\varphi) = 1-re^{i\varphi},
 & &
 e_2(\varphi) = 1+re^{i\varphi}.
\eq
We study this family for small positive $r$ and $\varphi \in [0,2\pi]$.
We consider the periods
\bq
 P_1(\varphi) \; = \; \int\limits_{\delta_1} \frac{dx}{y},
 & &
 P_2(\varphi) \; = \; \int\limits_{\delta_2} \frac{dx}{y},
 \;\;\;\;\;\;\;\;\;
 y \; = \; - \sqrt{x} \sqrt{x-e_1(\varphi)} \sqrt{x-e_2(\varphi)}, 
\eq
where the two cycles $\delta_1$ and $\delta_2$ form a basis of $H_1(E_\varphi,{\mathbb Z})$.
The orientation of the two cycles is such that
for $\varphi=0$ we have
\bq
 P_1(0) \; = \; 2 \int\limits_{0}^{e_1(0)} \frac{dx}{y}
 \; = \; - 2 \int\limits_{e_2(0)}^{\infty} \frac{dx}{y},
 & &
 P_2(0) \; = \; 2 \int\limits_{e_2(0)}^{e_1(0)} \frac{dx}{y},
\eq
where the path of integration in $x$-space is such that we have an infinitesimal small negative imaginary part for $x$.
We note that
\bq
 p(x,0) \;\; = \;\; p(x,\pi) \;\; = \;\; p(x,2\pi),
\eq
and
\bq
 e_1(0) \; = \; e_2(\pi) \; = \; e_1(2\pi),
 & &
 e_2(0) \; = \; e_1(\pi) \; = \; e_2(2\pi).
\eq
Thus, under a half-turn the polynomial $p(x,\varphi)$ transforms into itself, however the roots
$e_1(\varphi)$ and $e_2(\varphi)$ exchange their roles.
Let us now discuss the effect of a half-turn on the periods $P_1(\varphi)$ and $P_2(\varphi)$.
As $p(x,\varphi)$ transforms into itself, we may express the periods $P_1(\pi)$ and $P_2(\pi)$
as a linear combination of the periods $P_1(0)$ and $P_2(0)$.
\begin{figure}
\begin{center}
\includegraphics[scale=0.55]{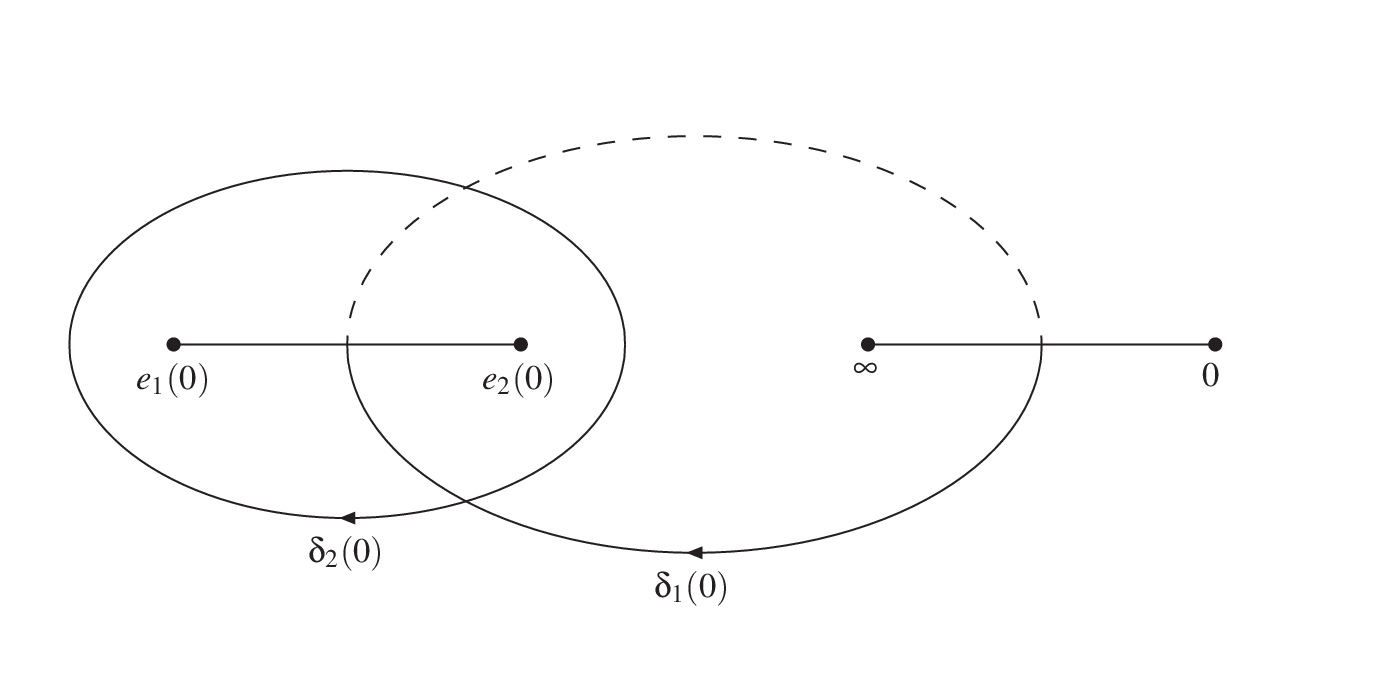}
\hspace*{5mm}
\includegraphics[scale=0.55]{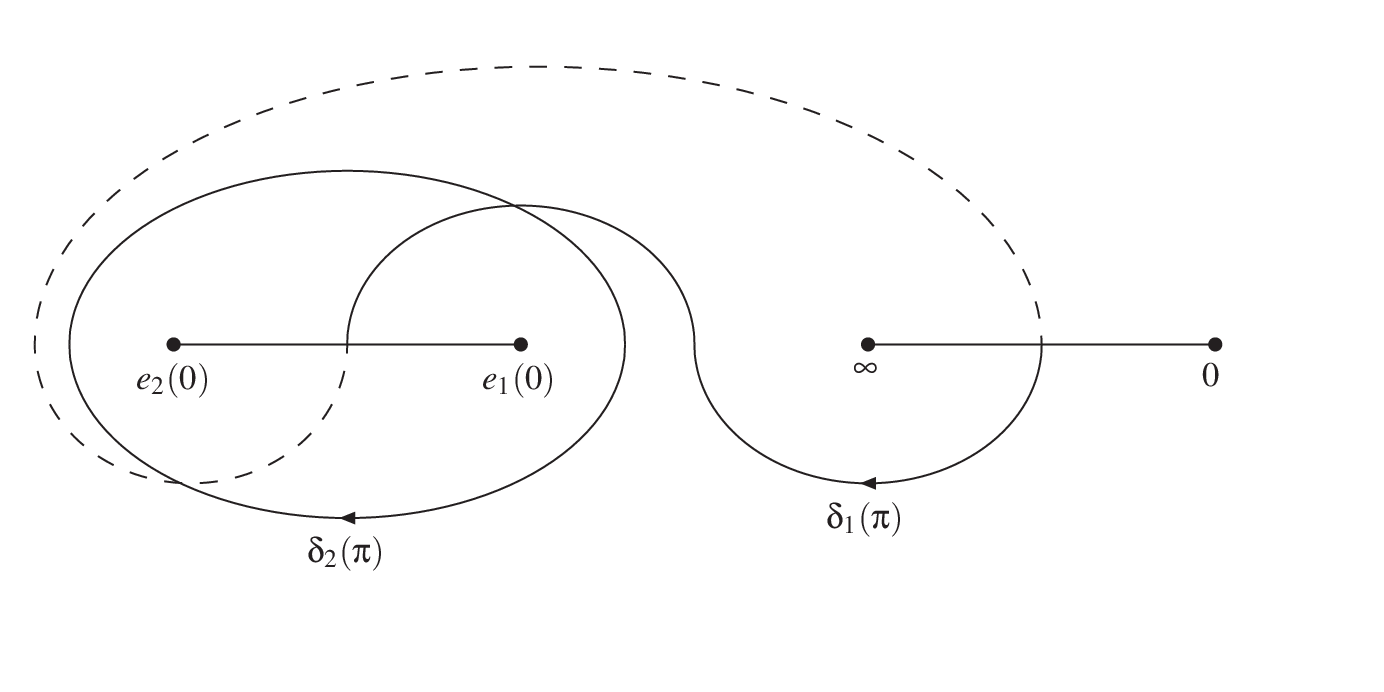}
\end{center}
\caption{\label{fig_period_path}
The left picture shows the cycles $\delta_1$ and $\delta_2$ for $\varphi=0$,
the right picture shows the cycles $\delta_1$ and $\delta_2$ for $\varphi=\pi$.
Solid lines corresponds to a path on one Riemann sheet, dashed lines to a path on the other Riemann sheet.
}
\end{figure}
As $\varphi$ ranges over the interval $[0,\pi]$ the cycles of integration for the two periods transform as shown in fig.~(\ref{fig_period_path}).
We recall that we may think of an elliptic curve as two copies of an Riemann sphere, each sphere with two cuts.
The elliptic curve is obtained by gluing together the two spheres at the cuts with opposite orientations.
From fig.~(\ref{fig_period_path}) it is clear that
\bq
 P_2(\pi) & = & P_2(0).
\eq
We may deform the cycle of integration $\delta_1(\pi)$ for $P_1(\pi)$ as shown in fig.~(\ref{fig_deformation_delta1}).
Thus
\bq
\label{monodromy_relation_P}
 P_1(\pi) & = & P_1(0) - P_2(0).
\eq
Eq.~(\ref{monodromy_relation_P}) is an application of the Picard-Lefschetz theorem.
In the example above $\delta_2$ is a vanishing cycle.
The Picard-Lefschetz theorem states that
\bq
 \delta_1\left(\pi\right)
 & = &
 \delta_1\left(0\right) \; - \; \left( \delta_1\left(0\right) \cdot \delta_2\left(0\right) \right) \; \delta_2\left(0\right),
\eq
where $(\delta_a \cdot \delta_b)$ denotes the intersection number of the cycles $\delta_a$ and $\delta_b$ \cite{Carlson}.

Combining two half-turns we obtain the monodromy relation
of eq.~(\ref{def_periods_final}) and eq.~(\ref{monodromy})
from
\bq
 P_1\left(2\pi\right) = P_1\left(0\right) - 2 P_2\left(0\right),
 & &
 P_2\left(2\pi\right) = P_2\left(0\right).
\eq


\section{Numerical results}
\label{sect:numerical_results}

In this section we consider three examples of Feynman integrals from the family of kite integrals.
The family of kite integrals 
\begin{figure}
\begin{center}
\includegraphics[scale=1.0]{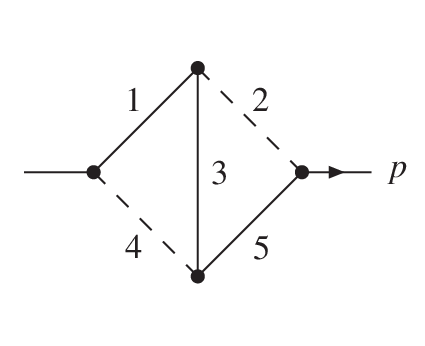}
\end{center}
\caption{
\it The kite graph. Solid lines correspond to massive propagators of mass $m$, dashed lines correspond to massless propagators.
}
\label{fig_kite_graph}
\end{figure}
is given in $D$-dimensional Minkowski space by 
\bq
\label{def_kite}
 I_{\nu_1 \nu_2 \nu_3 \nu_4 \nu_5}\left( D, p^2, m^2, \mu^2 \right)
 =
 \left(-1\right)^{\nu_{12345}}
 \left(\mu^2\right)^{\nu_{12345}-D}
 \int \frac{d^Dk_1}{i \pi^{\frac{D}{2}}} \frac{d^Dk_2}{i \pi^{\frac{D}{2}}}
 \frac{1}{D_1^{\nu_1} D_2^{\nu_2} D_3^{\nu_3} D_4^{\nu_4} D_5^{\nu_5}},
\eq
with the propagators
\bq
 D_1=k_1^2-m^2, \hspace{0.2cm}  
 D_2=k_2^2, \hspace{0.2cm}  
 D_3 = (k_1-k_2)^2-m^2, \hspace{0.2cm} 
 D_4=(k_1-p)^2, \hspace{0.2cm}  
 D_5 = (k_2-p)^2-m^2
\eq
and $\nu_{12345}=\nu_1+\nu_2+\nu_3+\nu_4+\nu_5$.
The kite graph is shown in fig.~(\ref{fig_kite_graph}).
The internal momenta are denoted by $k_1$ and $k_2$, 
the internal mass by $m$ and the external momentum by $p$. 
We denote $t=p^2$.
In the following we will suppress the dependence of the integrals on the mass $m$ and the scale $\mu$
and we write
\bq
 I_{\nu_1 \nu_2 \nu_3 \nu_4 \nu_5}(D,t) 
 & = & 
 I_{\nu_1 \nu_2 \nu_3 \nu_4 \nu_5}(D,t,m^2,\mu^2).
\eq
The three examples which we consider in this section are the kite integral
$I_{11111}\left(4-2\eps,t\right)$,
the equal-mass sunrise integral in $2-2\eps$ space-time dimensions
$I_{10101}\left(2-2\eps,t\right)$
and the integral (``bubble squared'')
$I_{21012}\left(4-2\eps,t\right)$.
\begin{figure}
\begin{center}
\includegraphics[scale=1.0]{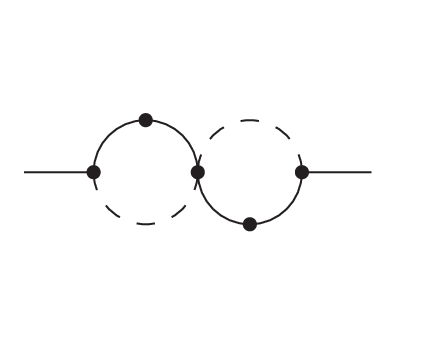}
\includegraphics[scale=1.0]{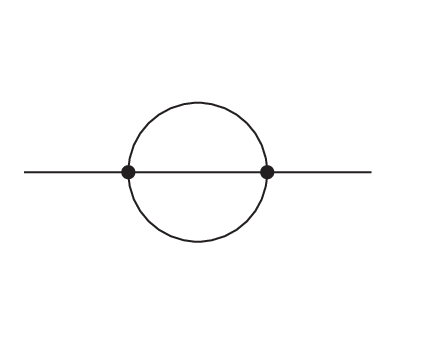}
\includegraphics[scale=1.0]{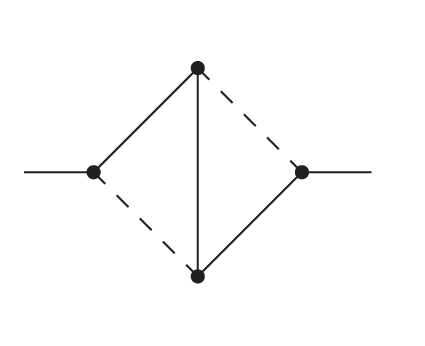}
\end{center}
\caption{
\it The Feynman graphs of the examples considered in this section.
A dot on a propagator indicates, that this propagator is raised to the power two.
}
\label{fig_Feynman_graphs}
\end{figure}
The Feynman graphs for these integrals are shown in fig.~(\ref{fig_Feynman_graphs}).
All three Feynman integrals start in the $\eps$-expansion at ${\mathcal O}(\eps^0)$.
In \cite{Adams:2016xah} we presented an algorithm to express all integrals from the kite family 
to all orders in the dimensional regularisation parameter in terms of $\mathrm{ELi}$-functions, which are
generalisations of (multiple) polylogarithms towards the elliptic case.
For the kite integral and the sunrise integral the occurrence of $\mathrm{ELi}$-functions is generic, 
the bubble integral squared may be expressed in terms of harmonic polylogarithms \cite{Vermaseren:1998uu,Remiddi:1999ew}.
All polylogarithms occurring in the integrals from the kite family may be expressed in terms of $\mathrm{ELi}$-functions. 
We include the trivial example of the bubble integral squared to show explicitly that our method includes the case where the $\mathrm{ELi}$-functions
degenerate to multiple polylogarithms.

Expressed in terms of $\mathrm{ELi}$-functions we have
\bq
\label{result_examples}
 I_{21012}\left(4-2\eps,t\right)
 & = &
 \left[ \frac{\mu^2}{t} \ln\left(1-\frac{t}{m^2}\right) \right]^2
 + {\mathcal O}\left(\varepsilon\right)
 \\
 & = &
 9 \frac{\mu^4}{t^2}
 \left[ 
          \overline{\mathrm{E}}_{1;0}\left( -1;  1; -q \right)
        - \overline{\mathrm{E}}_{1;0}\left( r_6; 1; -q \right)
 \right]^2
  + {\mathcal O}\left(\varepsilon\right),
 \nonumber \\
 I_{10101}\left(2-2\eps,t\right)
 & = &
 \frac{3 \psi_1}{\pi}
 \left[
 \mathrm{Cl}_2\left(\frac{2\pi}{3}\right)
 + \overline{\mathrm{E}}_{2;0}\left(r_3;-1;-q\right)
 \right]
 + {\mathcal O}\left(\varepsilon\right),
 \nonumber \\
 I_{11111}\left(4-2\eps,t\right)
 & = &
 \frac{\mu^2}{t}
 \left[
  2 G\left(0,1,1;y\right) - G\left(1,0,1;y\right)
  + \frac{\pi^2}{6} G\left(1;y\right)
 \right. \nonumber \\
 & & \left.
 + 27 \mathrm{Cl}_2\left(\frac{2\pi}{3}\right) \overline{\mathrm{E}}_{1;-1}\left(r_3;1;-q\right)
 + 27 \overline{\mathrm{E}}_{0,2;-2,0;2}\left(r_3,r_3;1,-1;-q\right)
 \right]
 + {\mathcal O}\left(\varepsilon\right).
 \nonumber
\eq
The notation is as follows: We introduced the dimensionless variable $y=t/m^2$. 
The symbol $r_n$ denotes the $n$-th root of unity
\bq
 r_n & = & \exp\left(\frac{2\pi i}{n} \right).
\eq
The $\overline{\mathrm{E}}$-functions are linear combinations of the $\mathrm{ELi}$-functions.
Both are defined in appendix~\ref{sec:ELi}.
$\mathrm{Cl}_2$ denotes the Clausen function, defined by
\bq
 \mathrm{Cl}_2\left(\varphi\right)
 & = &
 \frac{1}{2i} \left[ \mathrm{Li}_2\left(e^{i\varphi}\right) - \mathrm{Li}_2\left(e^{-i\varphi}\right) \right].
\eq
The harmonic polylogarithms $G(1;y)$, $G(0,1,1;y)$ and $G(1,0,1;y)$ may be expressed in terms of $\mathrm{ELi}$-functions.
We have
\bq
\lefteqn{
 G\left(1;y\right)
 =
 3 
 \left[ 
          \overline{\mathrm{E}}_{1;0}\left( -1;  1; -q \right)
        - \overline{\mathrm{E}}_{1;0}\left( r_6; 1; -q \right)
 \right],
 } & & 
 \\
\lefteqn{
 G\left(0,1,1;y\right)
 =
 9
 \left[ 
          \overline{\mathrm{E}}_{0,1;-1,0;4}\left( -1,-1;   1,1; -q \right)
        - \overline{\mathrm{E}}_{0,1;-1,0;4}\left( -1,r_6;  1,1; -q \right)
 \right. } & & \nonumber \\
 & & \left.
        - \overline{\mathrm{E}}_{0,1;-1,0;4}\left( r_6,-1;  1,1; -q \right)
        + \overline{\mathrm{E}}_{0,1;-1,0;4}\left( r_6,r_6; 1,1; -q \right)
 \right]
 \nonumber \\
 & &
 - 36
 \left[ 
          \overline{\mathrm{E}}_{0,0,1;-1,-1,0;2,2}\left( r_3,-1,-1;   -1,1,1; -q \right)
        - \overline{\mathrm{E}}_{0,0,1;-1,-1,0;2,2}\left( r_3,-1,r_6;  -1,1,1; -q \right)
 \right. \nonumber \\
 & & \left.
        - \overline{\mathrm{E}}_{0,0,1;-1,-1,0;2,2}\left( r_3,r_6,-1;  -1,1,1; -q \right)
        + \overline{\mathrm{E}}_{0,0,1;-1,-1,0;2,2}\left( r_3,r_6,r_6; -1,1,1; -q \right)
 \right],
 \nonumber \\
\lefteqn{
 G\left(1,0,1;y\right)
 =
 9
 \left[ 
          \overline{\mathrm{E}}_{0,2;-1,1;2}\left( -1,-1;   1,1; -q \right)
        - \overline{\mathrm{E}}_{0,2;-1,1;2}\left( -1,r_6;  1,1; -q \right)
 \right. } \nonumber \\
 & & \left.
        - \overline{\mathrm{E}}_{0,2;-1,1;2}\left( r_6,-1;  1,1; -q \right)
        + \overline{\mathrm{E}}_{0,2;-1,1;2}\left( r_6,r_6; 1,1; -q \right)
 \right]
 \nonumber \\
 & &
 - 36
 \left[ 
          \overline{\mathrm{E}}_{0,0,1;-1,-1,0;2,2}\left( -1,r_3,-1;  1,-1,1; -q \right)
        - \overline{\mathrm{E}}_{0,0,1;-1,-1,0;2,2}\left( -1,r_3,r_6; 1,-1,1; -q \right)
 \right. \nonumber \\
 & & \left.
        - \overline{\mathrm{E}}_{0,0,1;-1,-1,0;2,2}\left( r_6,r_3,-1;  1,-1,1; -q \right)
        + \overline{\mathrm{E}}_{0,0,1;-1,-1,0;2,2}\left( r_6,r_3,r_6; 1,-1,1; -q \right)
 \right].
 \nonumber
\eq
In the following we set $\mu=m$, express all integrals entirely in terms of $\mathrm{ELi}$-functions
and plot the real and the imaginary part of
the $\eps^0$-term of
\bq
 I_{21012}\left(4-2\eps,t\right),
 \;\;\;
 I_{10101}\left(2-2\eps,t\right),
 \;\;\;
 I_{11111}\left(4-2\eps,t\right)
\eq
as a function of $y=t/m^2$.
Note that the expressions in eq.~(\ref{result_examples}) are valid over the complete kinematic range $t \in {\mathbb R}$.
The periods $\psi_1$ and $\psi_2$ (and in turn the nome $q$) are computed according to eq.~(\ref{def_periods_final}).
We compare our results to numerical results obtained from the {\tt SecDec}-program \cite{Borowka:2015mxa,Borowka:2012yc,Borowka:2013cma,Borowka:2014aaa}.
\begin{figure}
\begin{center}
\includegraphics[scale=0.6]{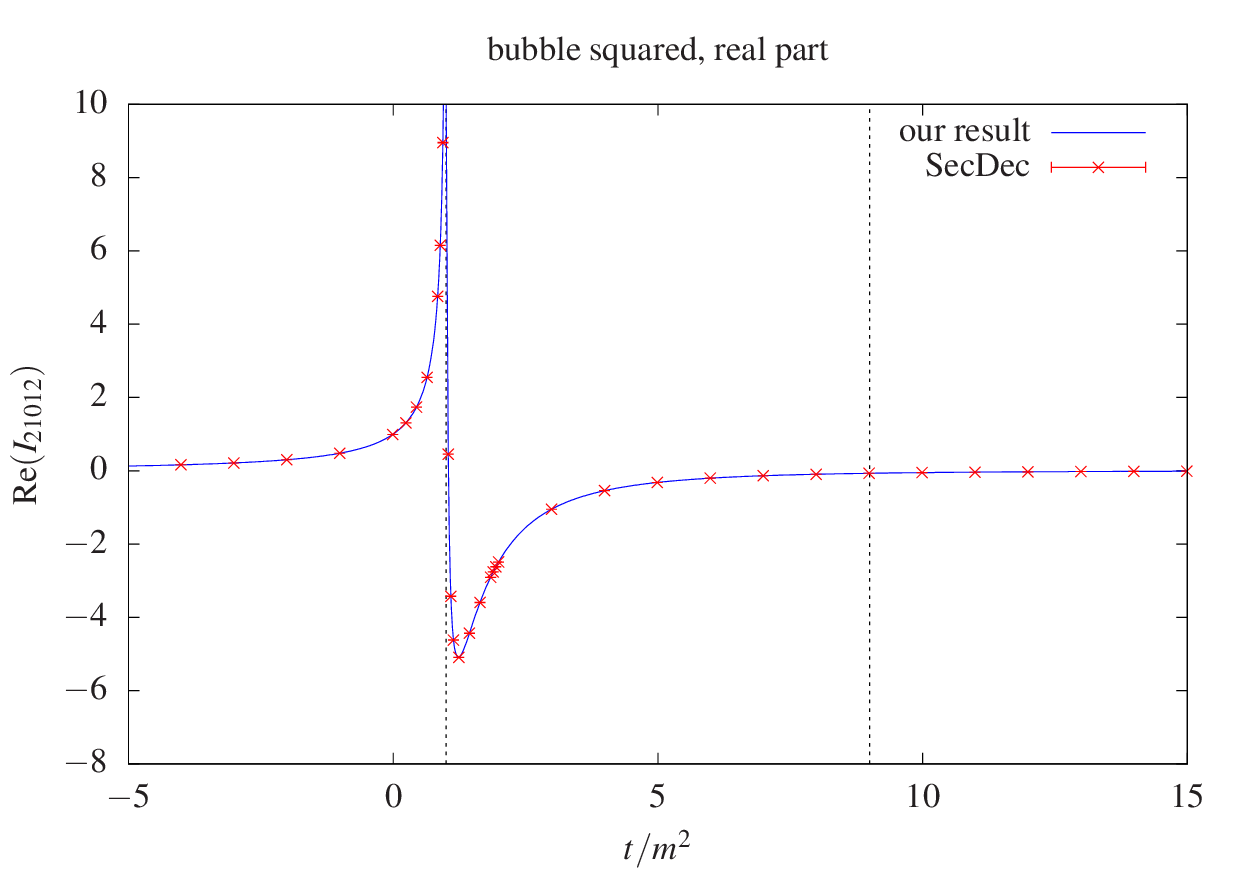}
\includegraphics[scale=0.6]{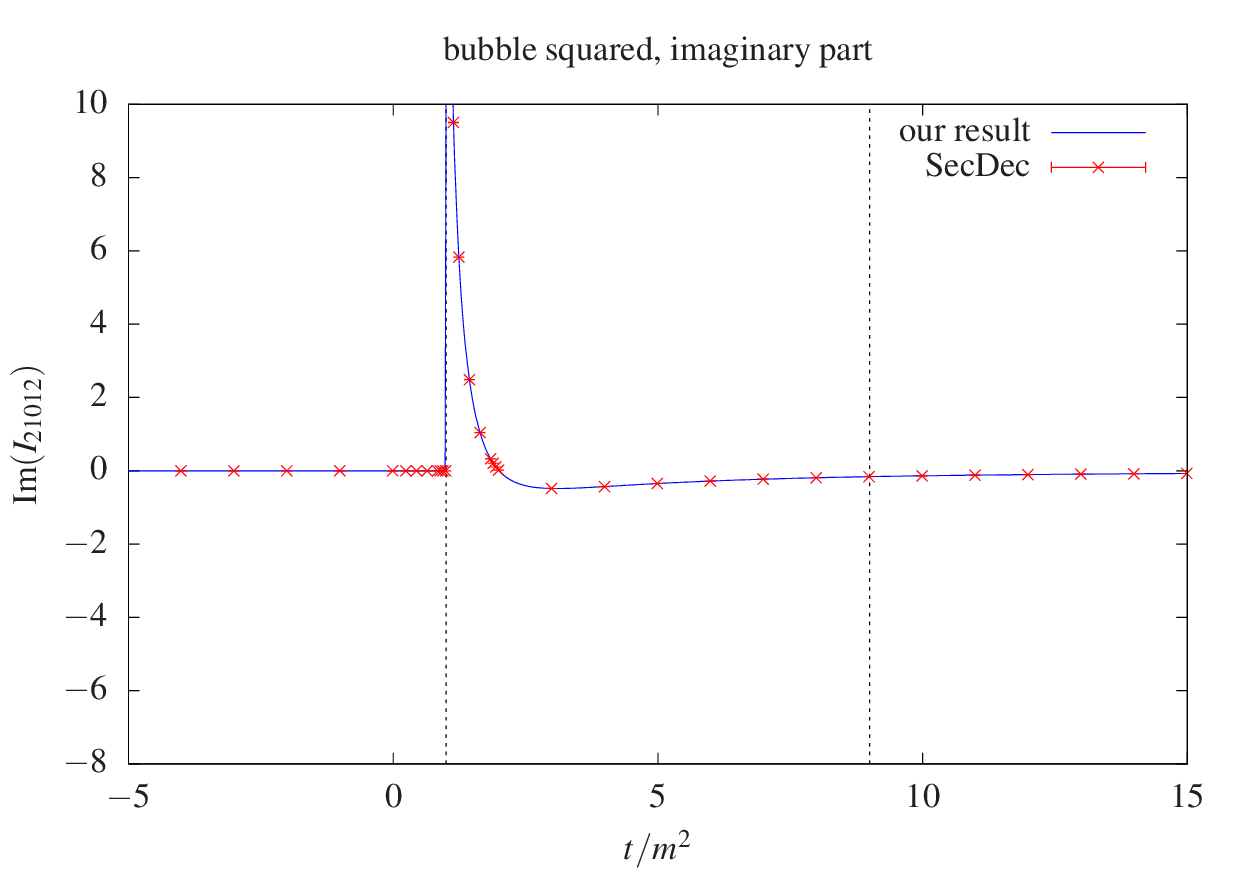}
\includegraphics[scale=0.6]{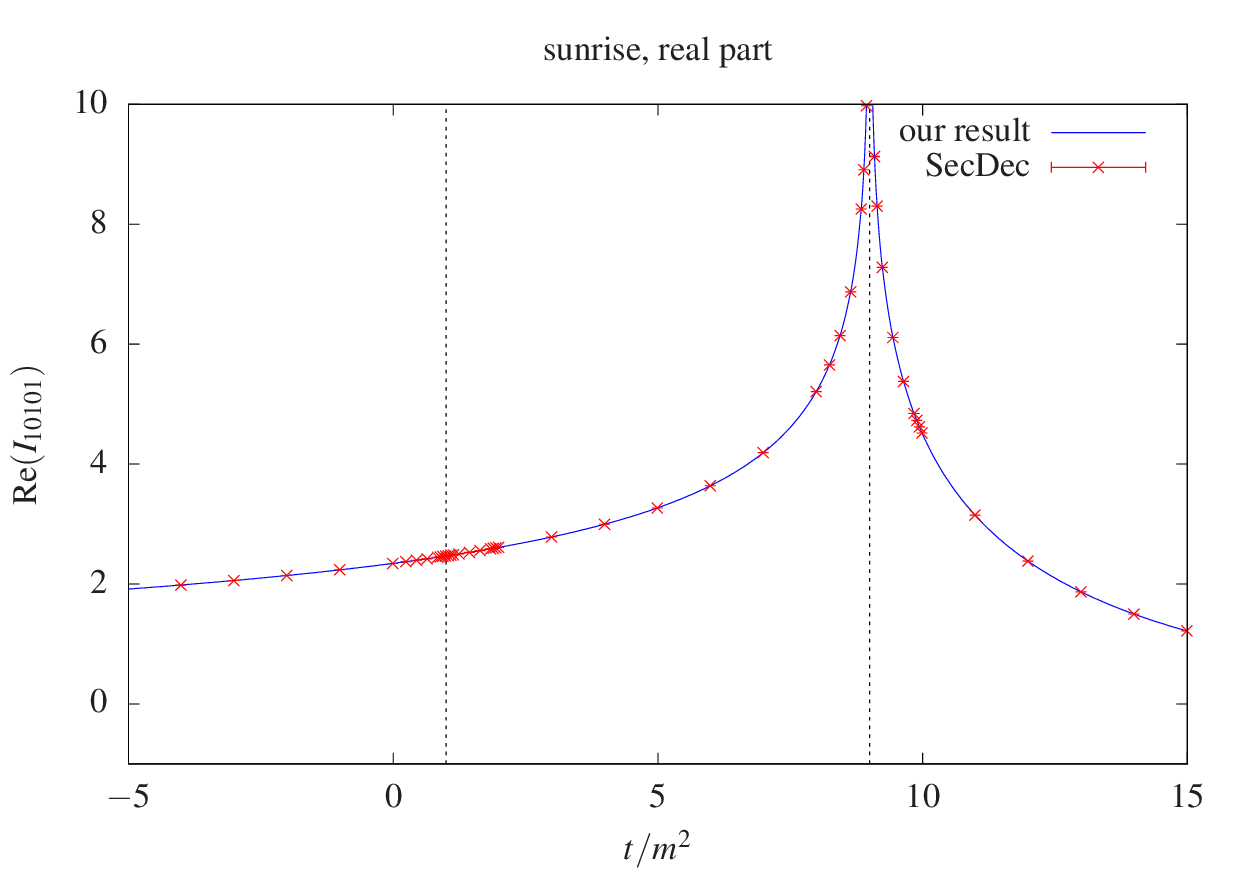}
\includegraphics[scale=0.6]{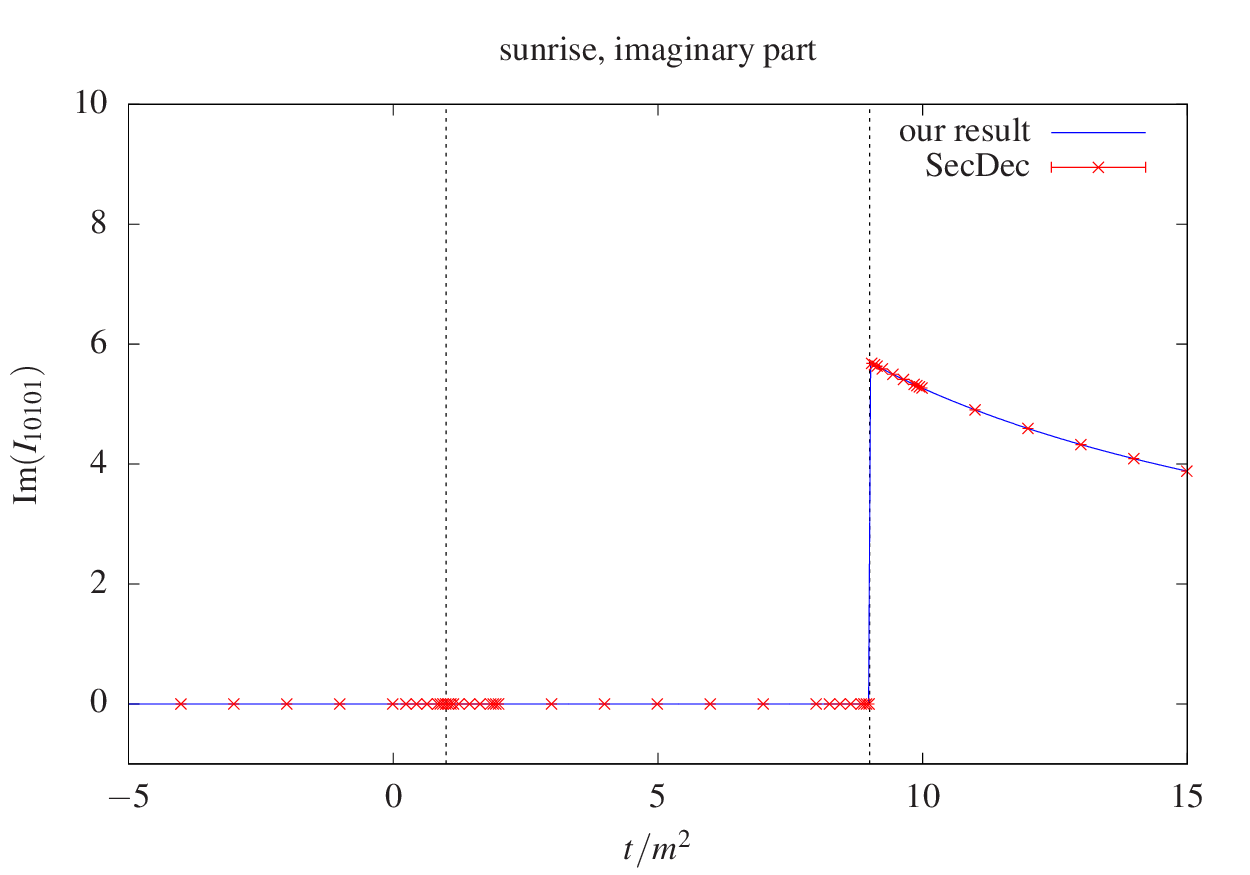}
\includegraphics[scale=0.6]{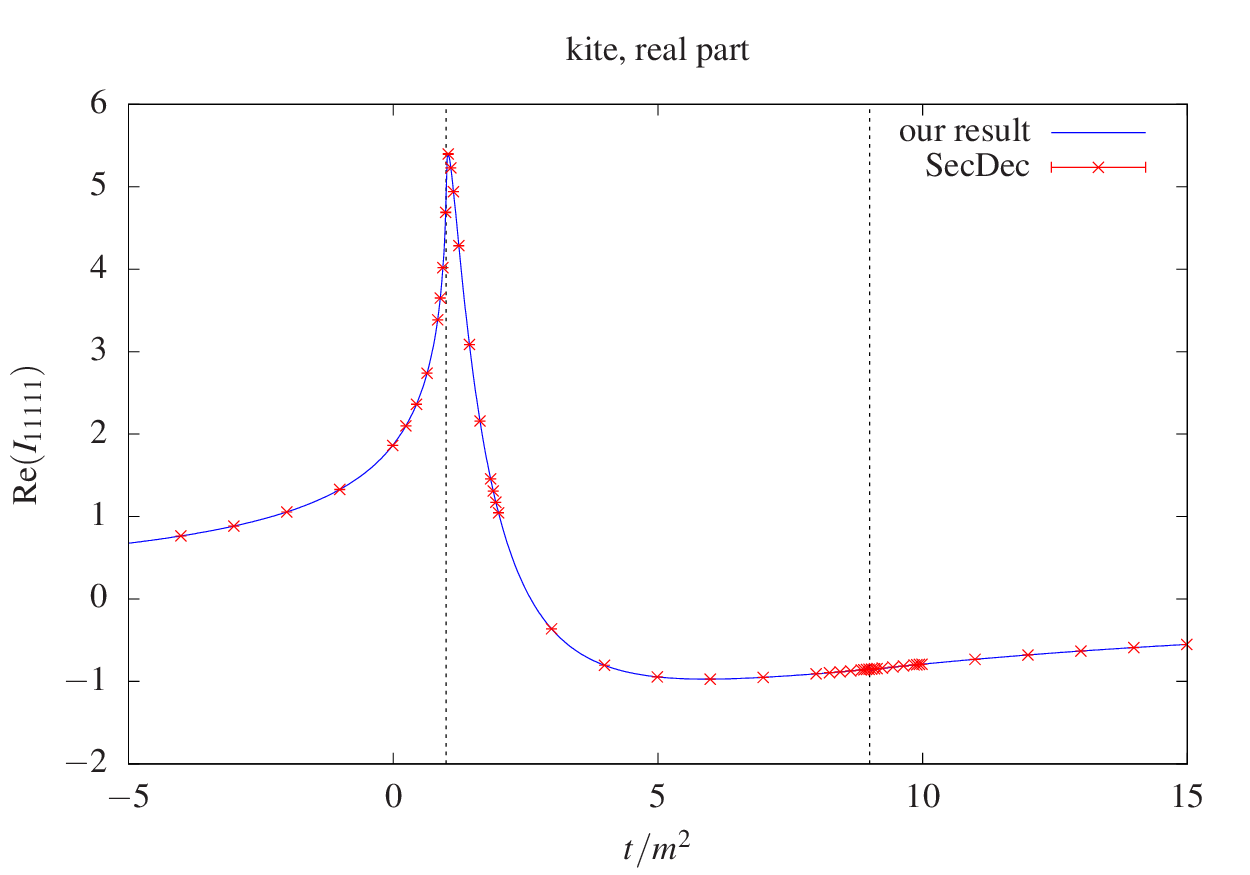}
\includegraphics[scale=0.6]{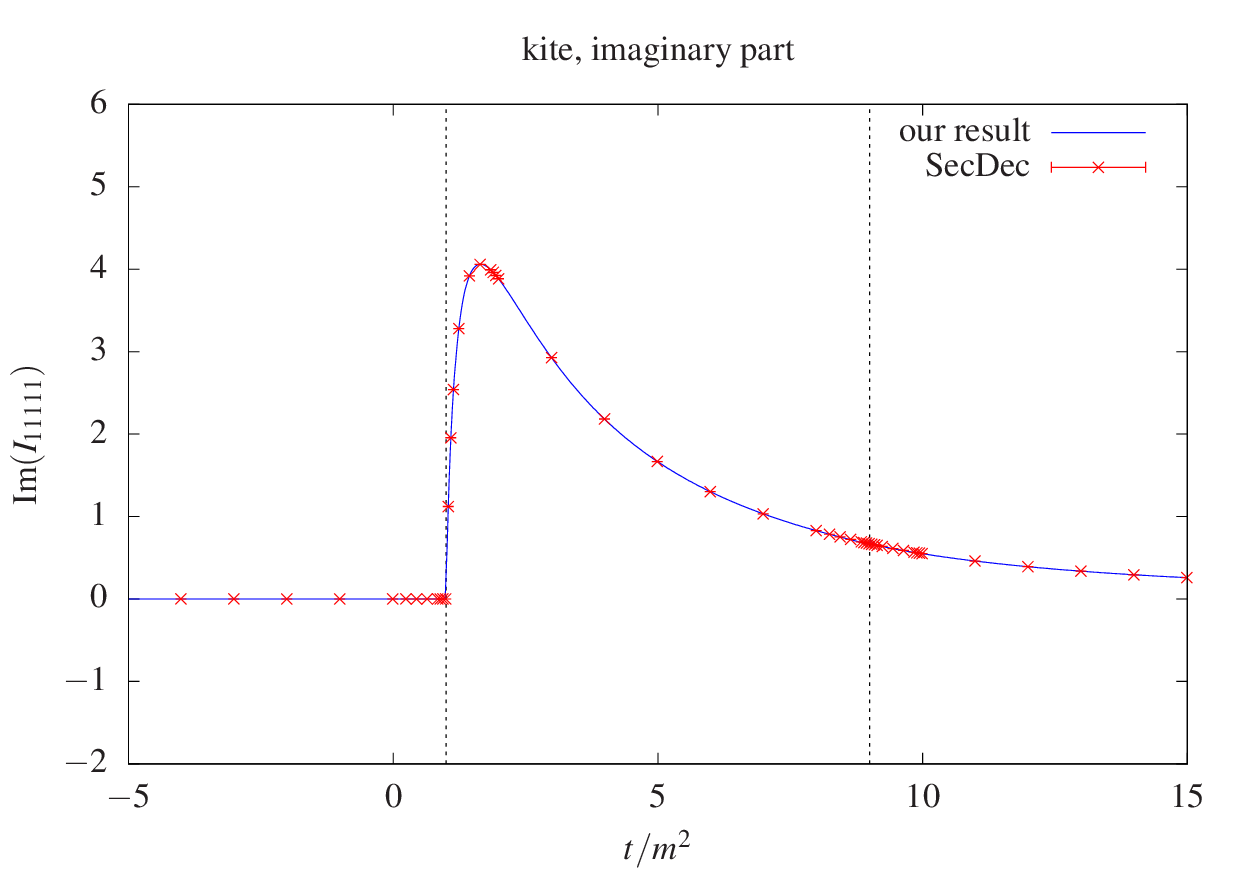}
\end{center}
\caption{\label{fig_result}
The real and the imaginary parts of the $\eps^0$-term of the Feynman integrals
$I_{21012}\left(4-2\eps,t\right)$,
$I_{10101}\left(2-2\eps,t\right)$ 
and
$I_{11111}\left(4-2\eps,t\right)$
as $y=t/m^2$ ranges over the interval $[-5,15]$.
The dashed vertical lines indicate the thresholds at $t=m^2$ and $t=9m^2$.
}
\end{figure}
The comparison is shown in fig.~(\ref{fig_result}) for $y \in [-5,15]$.
This range includes in particular the thresholds $t=m^2$ and $t=9m^2$.
We find perfect agreement over the full range $y \in {\mathbb R}$.

The results for the bubble integral squared $I_{21012}$ (i.e. the first two plots in fig.~(\ref{fig_result}))
show that the numerical evaluation of this integral in terms of $\mathrm{ELi}$-functions
correctly reproduces the result of this integral, which alternatively may be expressed in terms of multiple polylogarithms. 

\begin{figure}
\begin{center}
\includegraphics[scale=0.6]{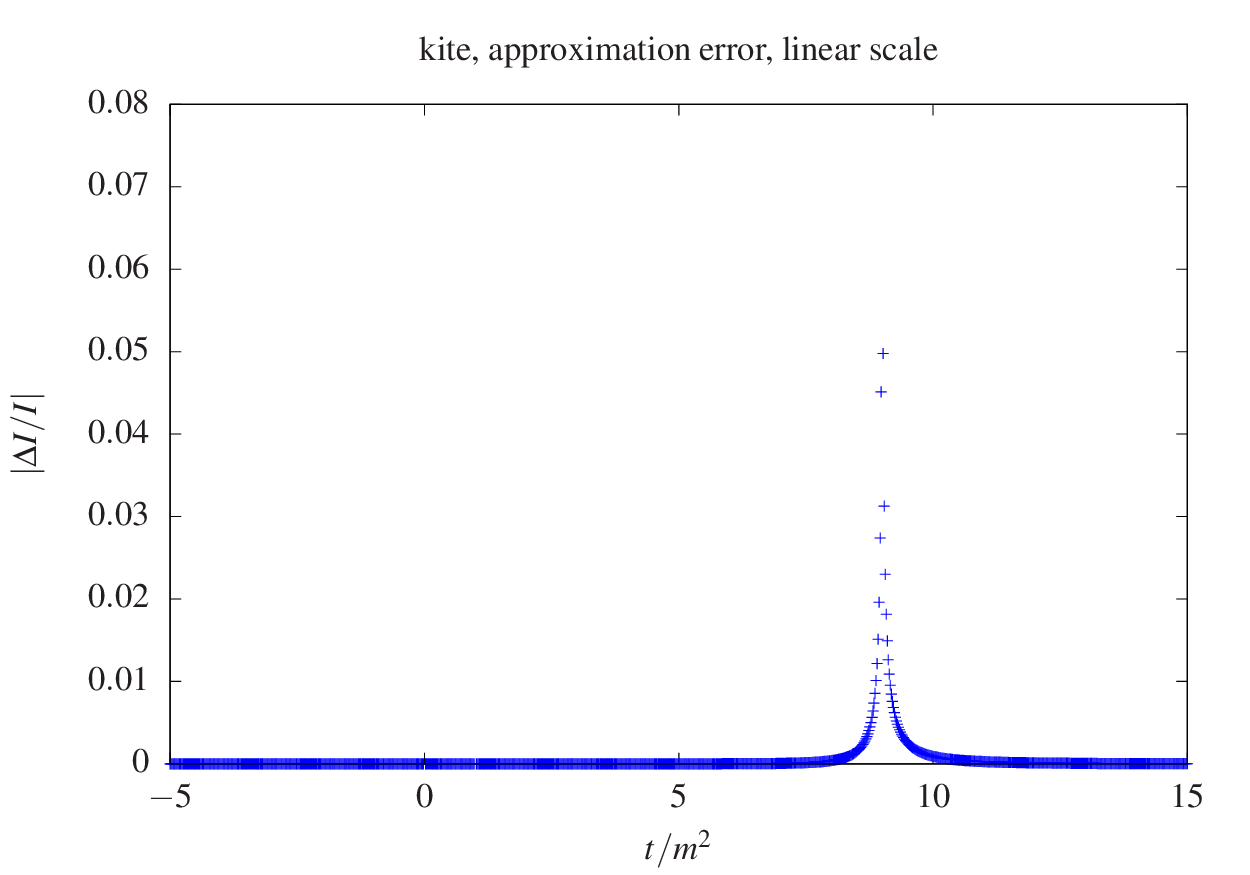}
\includegraphics[scale=0.6]{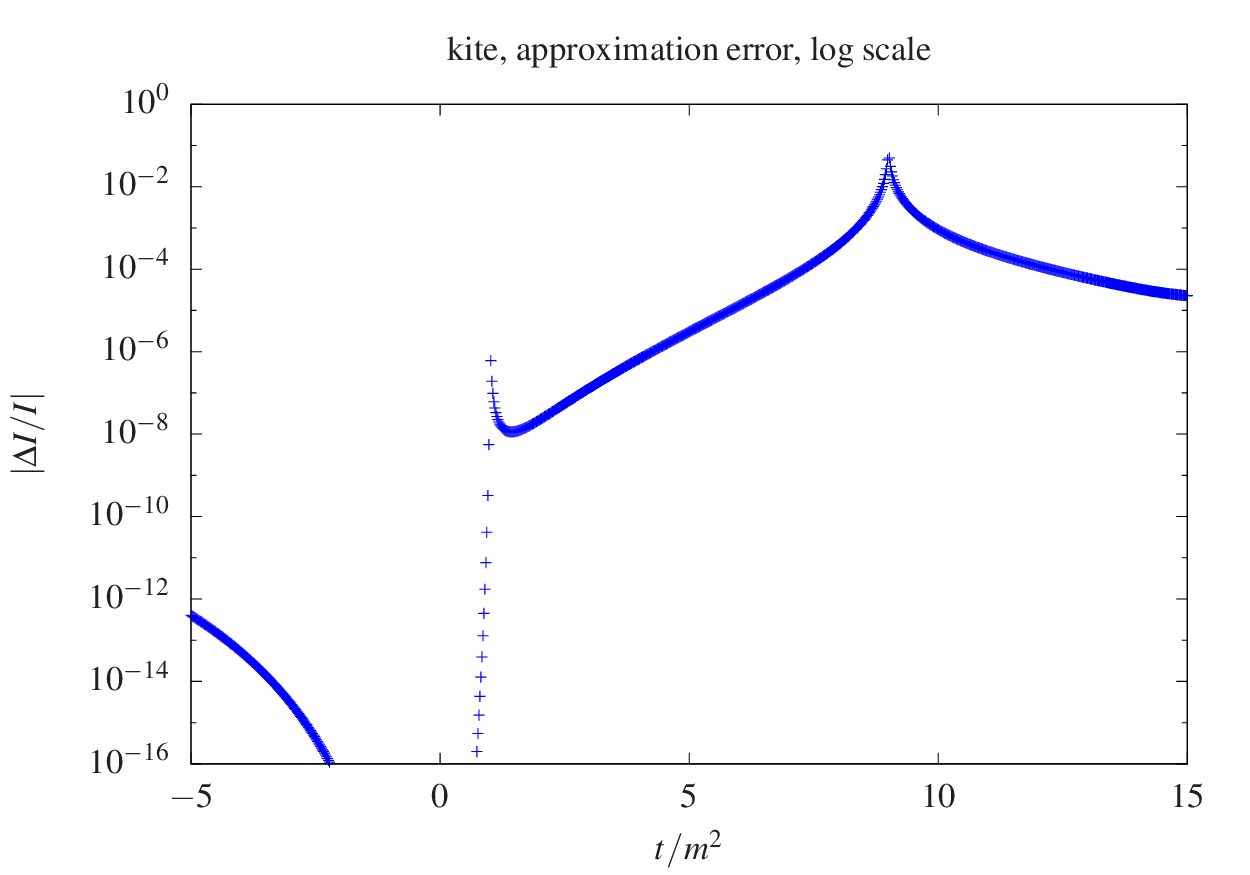}
\end{center}
\caption{\label{fig_result_error}
The relative difference of the ${\mathcal O}(q^{20})$-approximation with the ${\mathcal O}(q^{100})$-approximation
for the kite integral $I_{11111}$. The left plot shows the relative difference on a linear scale, the right plot
shows the relative difference on a logarithmic scale.
}
\end{figure}

We note that truncating the $q$-series at ${\mathcal O}(q^{20})$ gives already a very good approximation, including the thresholds.
In order to quantify this, we show in fig.~(\ref{fig_result_error}) for the kite integral
the relative difference of the ${\mathcal O}(q^{20})$-approximation with the ${\mathcal O}(q^{100})$-approximation,
defined by
\bq
 \left| \frac{\Delta I}{I} \right|
 & = &
 \left|
 \frac{ \left. I_{11111}\left(4,t\right) \right|_{q^{20}} - \left. I_{11111}\left(4,t\right) \right|_{q^{100}}}
      {\left. I_{11111}\left(4,t\right) \right|_{q^{100}}}
 \right|.
\eq
We evaluate the integral in the interval $t/m^2 \in [-5,15]$ at steps $\delta t / m^2 = 0.02$, excluding the singular points $t=m^2$ and
$t=9m^2$. Thus the closest points to the singular points are $t=0.98 m^2$, $t=1.02 m^2$, $t=8.98m^2$
and $9.02m^2$.
We see that the relative difference is rather small over a wide range of $t$, taking a maximum $|\Delta I | \approx 5 \%$
around $t=9m^2$, where the kite integral is continuous.
To see the behaviour at the threshold $t=m^2$, we show in the right plot of fig.~(\ref{fig_result_error})
the relative difference on a log-scale.
The relative difference around the threshold $t=m^2$ is $|\Delta I | \approx 10^{-6}$.
The integrals $I_{21012}$ and $I_{10101}$ show a similar behaviour. 
Again, the maximum of the relative difference for $t/m^2 \in [-5,15]$ is reached
around $t=9m^2$, being of the order of $1 \%$.
The $\mathrm{ELi}$-functions provide therefore a fast and efficient way to evaluate these integrals
over the complete kinematic range with the exception of small neighbourhoods 
around the singular points $t \in \{m^2, 9 m^2, \infty \}$,
where $|q|=1$.


\section{Conclusions}
\label{sect:conclusions}

In this paper we studied the analytic continuation and the numerical evaluation of Feynman integrals
from the kite family.
In previous publications we showed that these integrals can be expressed to all orders 
in the dimensional regularisation parameter $\eps$ in a neighbourhood of $t=0$ in terms of
elliptic generalisations of (multiple) polylogarithms, denoted as $\mathrm{ELi}$-functions.
In this paper we showed that these expressions hold in the full kinematic range $t \in {\mathbb R}$ after analytic
continuation of the two periods.
The $\mathrm{ELi}$-functions are power series in the variable $q$.
Furthermore we showed that $|q| \le 1$ holds in the full kinematic range $t \in {\mathbb R}$
and $|q|=1$ is attained only at the singular points $t\in\{m^2,9m^2,\infty\}$.
Therefore the $q$-series expansion of the $\mathrm{ELi}$-functions provide a fast and efficient way to evaluate these integrals
over a wide kinematic range. 
We compared results from this method to numerical results from the {\tt SecDec} program and found perfect agreement.

\subsection*{Acknowledgements}

C.B. and A.S. thank Konrad Schultka for useful discussions. 
S.W. would like to thank Hubert Spiesberger for useful discussions.
C.B. is supported by Deutsche Forschungsgemeinschaft under the project BO4500/1-1.

\begin{appendix}

\section{Convention regarding the roots of the cubic equation}
\label{sec:convention_roots}

In section~\ref{sec:elliptic_curve} we defined the roots $e_1$, $e_2$, $e_3$
of the cubic polynomial $4x^3-g_2x-g_3$ in eq.~(\ref{def_roots}).
The roots $e_1$ and $e_2$ were given by
\bq
\label{def_roots_recap}
 e_1 
 & = &
 \frac{1}{24 \mu^4} \left( -t^2 + 6 m^2 t + 3 m^4 + 3 \left( m^2 - t\right)^{\frac{3}{2}} \left( 9 m^2 - t \right)^{\frac{1}{2}} \right),
 \nonumber \\
 e_2 
 & = &
 \frac{1}{24 \mu^4} \left( -t^2 + 6 m^2 t + 3 m^4 - 3 \left( m^2 - t\right)^{\frac{3}{2}} \left( 9 m^2 - t \right)^{\frac{1}{2}} \right).
\eq
In this appendix we discuss the consequences of an alternative convention for the roots of the cubic equation, 
where $e_1$ and $e_2$ are replaced
by
\bq
\label{def_roots_alternative}
 \tilde{e}_1 
 & = &
 \frac{1}{24 \mu^4} \left( -t^2 + 6 m^2 t + 3 m^4 + 3 \sqrt{\left( m^2 - t\right)^3 \left( 9 m^2 - t \right)} \right),
 \nonumber \\
 \tilde{e}_2 
 & = &
 \frac{1}{24 \mu^4} \left( -t^2 + 6 m^2 t + 3 m^4 - 3 \sqrt{\left( m^2 - t\right)^3 \left( 9 m^2 - t \right)} \right),
\eq
respectively.

Let us first note that for $t<m^2$ and $9m^2<t$
one finds that  
$e_1$ agrees with $\tilde{e}_1$ and $e_2$ agrees with $\tilde{e}_2$.
Therefore the differences between the convention in eq.~(\ref{def_roots_recap})
and the one of eq.~(\ref{def_roots_alternative}) are restricted to the region
$m^2 < t < 9m^2$.
Let us first consider $e_1$ and $e_2$.
We recall that we assume that a infinitesimal small positive imaginary part is added to the variable $t$.
We have
\bq
 \left( m^2 - t - i0 \right)^{\frac{3}{2}}
 & = &
 i \left( t - m^2 \right)^{\frac{3}{2}}
 \;\;\;\;\;\; \mbox{for} \;\; t > m^2.
\eq
Thus
\bq
 \mathrm{Im}\left(e_1\right) \; > \; 0,
 \;\;\;
 \mathrm{Im}\left(e_2\right) \; < \; 0,
 \;\;\;\;\;\; \mbox{for} \;\; m^2 < t < 9 m^2.
\eq
Let us now consider $\tilde{e}_1$ and $\tilde{e}_2$.
Working to first order in the infinitesimal small imaginary part
one obtains
\bq
 \sqrt{\left( m^2 - t - i 0\right)^3 \left( 9 m^2 - t - i 0\right)}
 & = &
 \sqrt{\left( m^2 - t \right)^3 \left( 9 m^2 - t \right)  -4 \left( m^2 - t \right)^2 \left(7m^2-t\right) i 0 }.
 \nonumber
\eq
Thus
\bq
 \mathrm{Im}\left(\tilde{e}_1\right) 
 \;\;
 \left\{ \begin{array}{lr}
 < 0, & m^2 < t < 7 m^2, \\
 > 0, & 7 m^2 < t < 9 m^2, \\
 \end{array} \right.
 & &
 \mathrm{Im}\left(\tilde{e}_2\right) 
 \;\;
 \left\{ \begin{array}{lr}
 > 0, & m^2 < t < 7 m^2, \\
 < 0, & 7 m^2 < t < 9 m^2, \\
 \end{array} \right.
\eq
In the interval $m^2 < t < 7 m^2$ one has $\tilde{e_1}=e_2$ and $\tilde{e}_1=e_2$.
This just exchanges the roles of $e_1$ and $e_2$.
However, at $t=7m^2$ the roots $\tilde{e}_1$ and $\tilde{e}_2$ are not continuous functions of $t$.
For this reason it is better to use the definitions of the roots as given in eq.~(\ref{def_roots_recap}).

The discontinuity in $t$ at $7m^2$ has the following origin: For $7 m^2 < t < 9 m^2$ the imaginary part coming from
$(9m^2-t-i0)$ is larger and has opposite sign as the imaginary part coming from $(m^2-t-i0)$.

Similar considerations apply to the algebraic prefactor in eq.~(\ref{def_periods_II}), where we express
the periods $\psi_1$ and $\psi_2$ in terms of complete elliptic integrals of the first kind.
We defined the algebraic prefactor as
\bq
 \frac{4 \mu^2}{\left( m^2 - t\right)^{\frac{3}{4}} \left( 9 m^2 - t \right)^{\frac{1}{4}}},
\eq
and not as
\bq
 \frac{4 \mu^2}{\left[\left( m^2 - t\right)^3 \left( 9 m^2 - t \right)\right]^{\frac{1}{4}}}.
\eq
Both definitions agree for $t<m^2$, but the latter leads to spurious discontinuities in the analytic continuation.


\section{Elliptic generalisations of polylogarithms}
\label{sec:ELi}

In this appendix we define 
the $\mathrm{ELi}$-functions and the $\overline{\mathrm{E}}$-functions.
The latter are just linear combinations of the $\mathrm{ELi}$-functions.
Let us start with the $\mathrm{ELi}$-functions. These are 
functions of $(2l+1)$ variables $x_1$, ..., $x_l$, $y_1$, ..., $y_l$, $q$
and $(3l-1)$ indices $n_1$, ..., $n_l$, $m_1$, ..., $m_l$, $o_1$, ..., $o_{l-1}$.
For $l=1$ we set
\bq
 \mathrm{ELi}_{n;m}\left(x;y;q\right) 
 & = &
 \sum\limits_{j=1}^\infty \sum\limits_{k=1}^\infty \; \frac{x^j}{j^n} \frac{y^k}{k^m} q^{j k}.
\eq
For $l>1$ we define
\bq
\lefteqn{
 \mathrm{ELi}_{n_1,...,n_l;m_1,...,m_l;2o_1,...,2o_{l-1}}\left(x_1,...,x_l;y_1,...,y_l;q\right) 
 = }
 & & \nonumber \\
 & &
 \hspace*{15mm}
 = 
 \sum\limits_{j_1=1}^\infty ... \sum\limits_{j_l=1}^\infty
 \sum\limits_{k_1=1}^\infty ... \sum\limits_{k_l=1}^\infty
 \;\;
 \frac{x_1^{j_1}}{j_1^{n_1}} ... \frac{x_l^{j_l}}{j_l^{n_l}}
 \;\;
 \frac{y_1^{k_1}}{k_1^{m_1}} ... \frac{y_l^{k_l}}{k_l^{m_l}}
 \;\;
 \frac{q^{j_1 k_1 + ... + j_l k_l}}{\prod\limits_{i=1}^{l-1} \left(j_i k_i + ... + j_l k_l \right)^{o_i}}.
\eq
We have the relations
\bq
\label{ELi_multiplication}
\lefteqn{
 \mathrm{ELi}_{n_1;m_1}\left(x_1;y_1;q\right) 
 \mathrm{ELi}_{n_2,...,n_l;m_2,...,m_l;2o_2,...,2o_{l-1}}\left(x_2,...,x_l;y_2,...,y_l;q\right) 
 = } & & \nonumber \\
 & &
 \hspace*{25mm}
 =
 \mathrm{ELi}_{n_1,n_2,...,n_l;m_1,m_2,...,m_l;0,2o_2,...,2o_{l-1}}\left(x_1,x_2,...,x_l;y_1,y_2,...,y_l;q\right) 
\eq
and
\bq
\label{ELi_integration}
\lefteqn{
 \int\limits_0^{q} \frac{dq'}{q'}
 \mathrm{ELi}_{n_1,...,n_l;m_1,...,m_l;2o_1,2o_2,...,2o_{l-1}}\left(x_1,...,x_l;y_1,...,y_l;q'\right)
 = } & & \nonumber \\
 & &
 \hspace*{30mm}
 =
 \mathrm{ELi}_{n_1,...,n_l;m_1,...,m_l;2(o_1+1),2o_2,...,2o_{l-1}}\left(x_1,...,x_l;y_1,...,y_l;q\right).
\eq
It will be convenient to introduce abbreviations for certain linear combinations, which occur quite often.
We define a prefactor $c_n$ and a sign $s_n$, both depending on an index $n$ by
\bq
 c_n = \frac{1}{2} \left[ \left(1+i\right) + \left(1-i\right)\left(-1\right)^n\right] = 
 \left\{ \begin{array}{rl}
 1, & \mbox{$n$ even}, \\
 i, & \mbox{$n$ odd}, \\
 \end{array} \right.
 & &
 s_n = (-1)^n =
 \left\{ \begin{array}{rl}
 1, & \mbox{$n$ even}, \\
 -1, & \mbox{$n$ odd}. \\
 \end{array} \right.
\eq
For $l=1$ we define the linear combinations
\bq
 \overline{\mathrm{E}}_{n;m}\left(x;y;q\right) 
 & = &
 \frac{c_{n+m}}{i}
 \left[
  \mathrm{ELi}_{n;m}\left(x;y;q\right)
  - s_{n+m} \mathrm{ELi}_{n;m}\left(x^{-1};y^{-1};q\right)
 \right].
\eq
More explicitly, we have
\bq
\label{def_Ebar_weight_1}
 \overline{\mathrm{E}}_{n;m}\left(x;y;q\right) 
 & = &
 \left\{ \begin{array}{ll}
 \frac{1}{i}
 \left[
 \mathrm{ELi}_{n;m}\left(x;y;q\right) - \mathrm{ELi}_{n;m}\left(x^{-1};y^{-1};q\right)
 \right],
 & \mbox{$n+m$ even,} \\
 & \\
 \mathrm{ELi}_{n;m}\left(x;y;q\right) + \mathrm{ELi}_{n;m}\left(x^{-1};y^{-1};q\right),
 & \mbox{$n+m$ odd.} \\
 \end{array}
 \right.
\eq
For $l>0$ we proceed as follows:
For $o_1=0$ we set
\bq
\label{Ebar_multiplication}
\lefteqn{
 \overline{\mathrm{E}}_{n_1,...,n_l;m_1,...,m_l;0,2o_2,...,2o_{l-1}}\left(x_1,...,x_l;y_1,...,y_l;q\right) 
 = } & & \nonumber \\
 & &
 \hspace*{20mm}
 =
 \overline{\mathrm{E}}_{n_1;m_1}\left(x_1;y_1;q\right) 
 \overline{\mathrm{E}}_{n_2,...,n_l;m_2,...,m_l;2o_2,...,2o_{l-1}}\left(x_2,...,x_l;y_2,...,y_l;q\right).
\eq
For $o_1 > 0$ we set recursively
\bq
\label{Ebar_integration}
\lefteqn{
 \overline{\mathrm{E}}_{n_1,...,n_l;m_1,...,m_l;2o_1,2o_2,...,2o_{l-1}}\left(x_1,...,x_l;y_1,...,y_l;q\right) 
 = 
 } & & \nonumber \\
 & &
 \hspace*{20mm}
 =
 \int\limits_0^{q} \frac{dq'}{q'} 
 \overline{\mathrm{E}}_{n_1,...,n_l;m_1,...,m_l;2(o_1-1),2o_2,...,2o_{l-1}}\left(x_1,...,x_l;y_1,...,y_l;q'\right).
\eq
The $\overline{\mathrm{E}}$-functions are
linear combinations of the $\mathrm{ELi}$-functions with the same indices.
More concretely, an $\overline{\mathrm{E}}$-function of depth $l$ can be expressed
as a linear combination of $2^l$ $\mathrm{ELi}$-functions.
We have
\bq
\lefteqn{
 \overline{\mathrm{E}}_{n_1,...,n_l;m_1,...,m_l;2o_1,...,2o_{l-1}}\left(x_1,...,x_l;y_1,...,y_l;q\right) 
 = 
 } & & \\
 & &
 =
 \sum\limits_{t_1=0}^1 ... \sum\limits_{t_l=0}^1
 \left[ \prod\limits_{j=1}^l \frac{c_{n_j+m_j}}{i} \left( - s_{n_j+m_j} \right)^{t_j} \right]
 \mathrm{ELi}_{n_1,...,n_l;m_1,...,m_l;2o_1,...,2o_{l-1}}\left(x_1^{s_{t_1}},...,x_l^{s_{t_l}};y_1^{s_{t_1}},...,y_l^{s_{t_l}};q\right).
 \nonumber
\eq


\section{The arithmetic-geometric mean}
\label{sec:agm}

In this appendix we review the numerical evaluation of the complete elliptic integral of the first kind with the
help of the arithmetic-geometric mean.
Let $a_0$ and $b_0$ be two complex numbers.
For $n \in {\mathbb N}_0$ one sets
\bq
 a_{n+1} \; = \; \frac{1}{2} \left( a_n+b_n \right),
 & &
 b_{n+1} \; = \; \pm \sqrt{a_n b_n}.
\eq
The sign of the square root is chosen such that \cite{Cox:1984} 
\bq
 \left| a_{n+1} - b_{n+1} \right| & \le & \left| a_{n+1} + b_{n+1} \right|,
\eq
and in case of equality one demands in addition
\bq
 \mathrm{Im}\left( \frac{b_{n+1}}{a_{n+1}} \right) & > & 0.
\eq
The sequences $(a_n)$ and $(b_n)$ converge to a common limit 
\bq
 \lim\limits_{n \rightarrow \infty} a_n \; = \;
 \lim\limits_{n \rightarrow \infty} b_n
 \; = \; 
 \mathrm{agm}(a_0,b_0),
\eq
known as the arithmetic-geometric mean.
The complete elliptic integral of the first kind is given by
\bq
 K\left(k\right)
 & = &
 \frac{\pi}{2 \; \mathrm{agm}\left(k',1\right)},
 \;\;\;
 k' \; = \; \sqrt{1-k^2}.
\eq

\end{appendix}

\bibliography{/home/stefanw/notes/biblio}
\bibliographystyle{/home/stefanw/latex-style/h-physrev5}

\end{document}